\documentclass[a4paper,11pt]{article}
\usepackage[utf8]{inputenc}

\usepackage[a4paper,text={16.5cm,25.2cm},centering]{geometry}
\usepackage{lmodern}
\usepackage{bm}
\usepackage{graphicx}
\usepackage{microtype}
\usepackage{upquote}
\usepackage{listings}
\usepackage{xcolor}
\usepackage{array}

\usepackage{amsmath}
\usepackage{amssymb}
\usepackage{amsfonts}
\usepackage{color}
\usepackage{nicefrac}
\usepackage{multirow}

	\newcommand{\be}{\begin{equation}}
	\newcommand{\ee}{\end{equation}}
	\newcommand{\ben}{\begin{equation*}}
	\newcommand{\een}{\end{equation*}}

	\newcommand{\usrc}{u^{\text{src}}}
	
	\newcommand{\usrchsg}{\overline{u}^{\text{src}}}

	\newcommand{\psrc}{p^{\text{src}}}
	\newcommand{\muller}{\mathcal{M}}
	
	\newcommand{\vb}[1]{\ifmmode\mathbf{#1}\else$\mathbf{#1}$\fi} 

	\newcommand{\HSgreen}{\overline{G}}
	
	\newcommand{\argpot}[1]{ [\: #1 \:] }
	
	\newcommand{\vet}[1]{\overline{#1}}
 	\newcommand{\xsrc}{x_{\text{src}}}
 	\newcommand{\bfs}[1]{ \boldsymbol{ #1 } }
 	\newcommand{\TL}{ \text{TL} }

\title{A 3D acoustic propagation model for shallow waters based on an indirect Boundary Element Method}
\author{E. F. Lavia, J. D. Gonzalez \& S. Blanc}
\date{ \today }

\begin{document}

\maketitle

\begin{abstract}
The purpose of this work is twofold: (a) To present the theoretical formulation of a 3D acoustic propagation model based 
on a Boundary Element Method (BEM), which uses a half-space Green function in place of the more conventional free-space 
Green function and (b) to show a numerical implementation aimed to explore the formulation in simple idealized cases 
--controlled by a few parameters--, and provides necessary tests for the accuracy and performance of the model. The 
half-space Green's function, which has been used previously in scattering and diffraction, adds terms to the usual 
expressions of the integral operators without altering their continuity properties.
Verifications against the Pekeris waveguide suggest that the model allows an adequately prediction for the acoustic 
field. Likewise, numerical explorations in relation to the necessary mesh size for the description of the water-sediment 
interface allow us to conclude that a TL prediction with acceptable accuracy can be obtained with the use of a bounded 
mesh around the desired evaluation region.
\end{abstract}

\section{Introduction}
\label{intro}

Underwater sound propagation is significantly affected by three-dimensional
(3D) effects due to the complex realistic ocean environment.
The importance of these effects has been early recognised
\cite{buckingham1992ocean, tolstoy19963} and even emphasized
by at-sea and/or at-laboratory measurements \cite{sturm2013comparisons}.


A great variety of models have been developed to estimate underwater sound
propagation in 3D scenarios that, according to their governing equations and
numerical approaches, can be categorized into three main groups
\cite{oliveira2021underwater}, namely, (a) extended parabolic equation (PE)
models, such as the efficient marching solution based on the parabolic equation
proposed by \cite{lin2013three} that is applied to a local and to a global
ocean environment by \cite{heaney2016three} and \cite{lin2019three},
respectively; (b) normal mode models \cite{taroudakis1990variational,
porter1992kraken, decourcy2020coupled}; and (c) ray and beam tracing models
\cite{de2018simplex, porter2019beam}. However, modelling acoustic propagation
for 3D environments is still a significant two-fold challenge due to both the
difficulties associated with a thorough comprehension of the physical phenomenon
and the high computational time costs.

On the other hand, methods based on boundary integration, particularly the
Boundary Element Method (BEM), have been applied to a stratified oceanic media
in shallow waters. Although the BEM is computationally expensive than other
methods, acceleration algorithms can be used to improve its performance but it
is still not capable to model realistic scenarios for long-range propagation
(several thousand of kilometres).  Nevertheless, the method still has its
strong advantages such as a clean mathematical formulation and a consistent
behaviour at infinity (where the fields are expected to vanish).

The BEM consists on defining the acoustic field through surface-integral
operators reducing the definition of the computational domain from 3D to 2D.
Several references can be found in the literature, as \cite{li2019fast} who
used a multi-layer approach for shallow waters by defining boundary integral
equations whose unknowns are the pressure field and the normal particle velocity
on the surfaces; this approach can be categorized as a direct BEM approach.

The continuous growth in computational power (hardware equipped with more raw
processing power and more memory capacity) opens up the game for methods based
on volume or boundary discretization as the Finite Element Method (FEM) and the
Boundary Element Method (BEM) (these examples are not exhaustive). These methods
allows for a more realistic modelling since they assume fewer hypotheses than
classical and well established ones as the ray methods, parabolic equation
methods and coupled normal modes \cite{boyles1984acoustic}, which have been
successfully used for several decades. The above mentioned assumptions generally
refer to symmetry conditions or their validity in certain frequency regimes.

In this work we present a numerical implementation of BEM capable of tackling
the acoustic propagation problem in a homogeneous medium with an arbitrary
bathymetry (range-dependent model) that can be easily extended to a multilayer
coupled approach to consider inhomogeneous media. This formulation includes an
appropriate half-space Green function in order to take into account an infinite
pressure release surface which is a representation of the ocean free surface in
an exact analytical way. This idea has already been used in
\cite{dawson1990boundary} for a BIEM (Boundary Integral Equation Methods)
solution for the acoustic scattering from the surface of a waveguide then
applied to a two-dimensional case with non-penetrable boundaries, in
\cite{santiago2004modified},  and in \cite{seybert1988radiation} within the
context of acoustic scattering and radiation.
The approach presented here uses a modified green function in order to avoid
generating a mesh corresponding to the interface ocean-atmosphere. The indirect
BEM leads to operators that are numerically well-behaved in comparison with the
direct methods such as \cite{li2019fast} that needs some preconditioners to
numerically solve the problem.

This paper is organized as follows. In Section 2 the BEM model is presented. The
boundary value problem corresponding to acoustic propagation in a waveguide is
stated in Section 2.1, then the BEM theoretical formulation based on a
half-space Green function –emphasizing the differences with those based on the
free-space function– is developed in Section 2.2, and finally an outline of the
numerical method is provided in Section 2.3. Certain more technical mathematical
aspects and some intermediate steps in the derivations are confined to the
appendices. In Section 3 numerical experiments aided to verify the BEM model are
carried out. To this end a wave-number integration technique applied to the
propagation problem of a Pekeris waveguide provided with a sediment layer (with
and without attenuation) is used as a benchmark.  Section 4 shows the
performance of the model in a shallow water waveguide with a conical
mountain, which provides as a range-dependent bathymetry. The conclusions of the
work and suitable further steps aimed to improve the model are summarized in
Section 5.

\section{Model Formulation}
\label{model}

\subsection{Acoustic propagation problem}
\label{propagation_problem}

\begin{figure}[!h]
    \centering
	\includegraphics[scale=0.6]{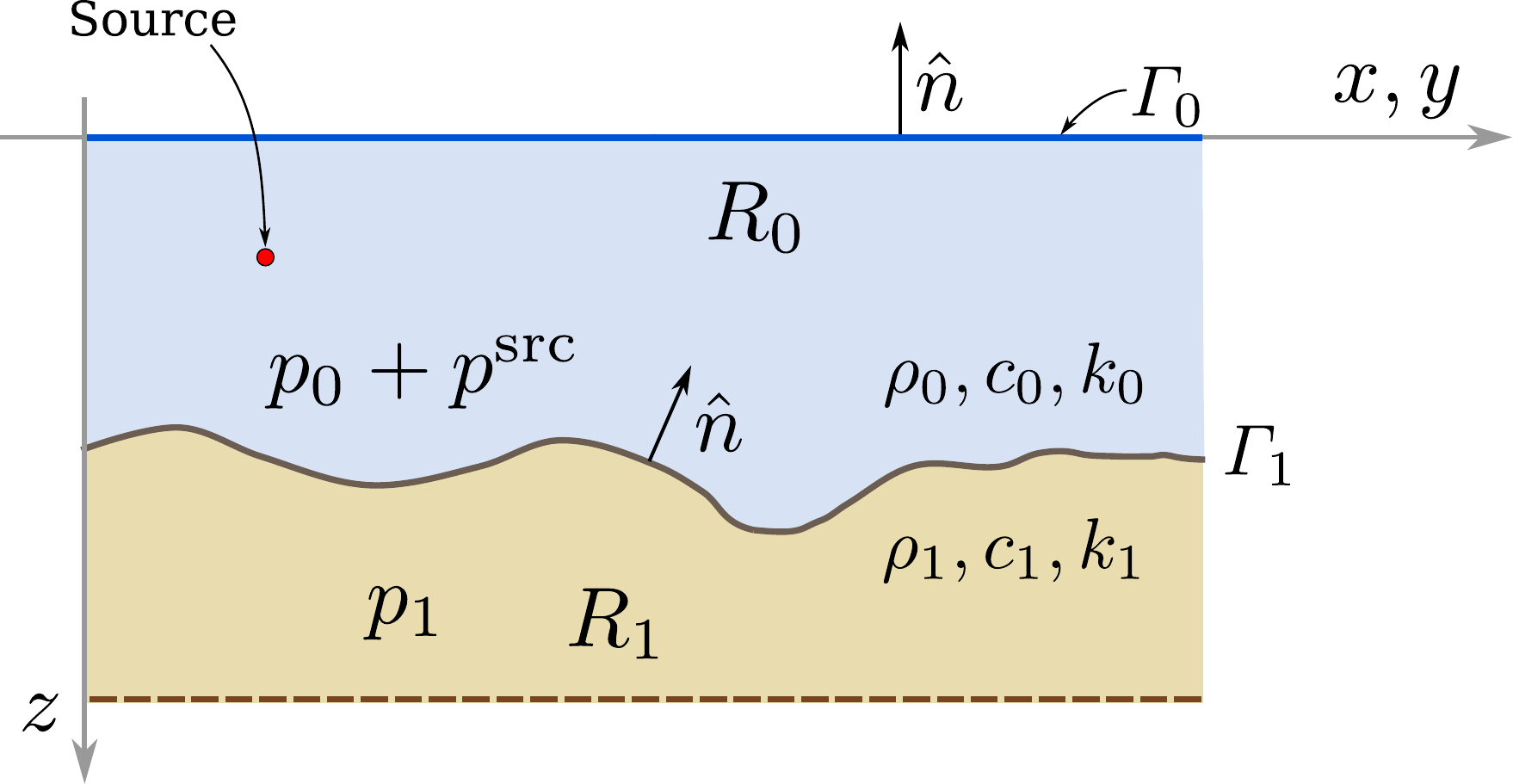}
	\caption{Scheme for the propagation problem.
	It should be noted that neither cylindrical nor plane symmetry must be inferred from this 2D sketch.}
	\label{scheme1}
\end{figure}

The problem of the underwater sound propagation in a homogeneous waveguide with range dependent bathymetry
is schematically shown in Figure \ref{scheme1}.
A time harmonic point source of circular frequency $\omega$ is located in an homogeneous water layer 
(which defines the volumetric region $R_0$) bounded above by the plane surface $\Gamma_0$ 
and below by the arbitrary surface $\Gamma_1$. The latter limits an unbounded volumetric region $R_1$
that constitutes a marine sediment layer.
The surfaces $\Gamma_0, \Gamma_1$ represent the interfaces seawater-atmosphere  and seawater-marine sediments,
respectively. Each region $R_i$ ($i=0,1$) is characterized by constant acoustic properties $c_i, \rho_i$ (sound
speed and density) and wavenumber $k_i = \omega /c_i$.
The boundless nature of the $R_1$ region is emphasized in the scheme by the dashed line at the bottom. 

The given point source generates, through interaction with the boundaries, acoustic pressure fields $p_i$ in each
region $R_i$ that are solutions of the scalar Helmholtz equation
\be
	( \: \nabla^2 + k_i^2 \:) \: p_i = 0.
	\label{helmh_eqs}
\ee
If the source field is $\psrc$, then the total field in the region $R_0$ is $p = p_0 + \psrc $.
As it is usual in propagation problems, {the boundary condition at} $\Gamma_0$ assumes a
pressure-release surface (Dirichlet) ($p_0 + \psrc = 0$), whereas at $\Gamma_1$ the pressure field must
satisfy the transmission condition that arises from demanding continuity of both,
the field and the normal particle velocity
$ v_n = - 1 / (i \omega \rho ) \partial_n p $ (the notation $\partial_n$ is a common abbreviation for 
the normal derivative $\hat{n}\cdot\nabla$).
The point source located at $\xsrc$ is assumed to generate a field 
\[
 	\psrc(x)= \frac{ e^{ i k_0 |x - \xsrc |}}{ |x - \xsrc| },
\]
with unitary amplitude.



According to the mathematical framework appropriate to BEM (integral operators theory), it is convenient
\cite{kress1978transmission} to transform the fields according to
\be
	u_0 = \frac{p_0}{\rho_0}, \qquad \qquad 
	u_1 = \frac{p_1}{\rho_1}, \qquad \qquad 
	\usrc = \frac{ \psrc }{\rho_0},
	\label{field_transform}
\ee
which has the consequence of making the factors $\rho_i^{-1}$ drop out from the normal derivatives.
In terms of these fields $u_i$ the pressure release condition at $\Gamma_0$ leads to
\be
	u_0(x) + \usrc(x) =\: 0 \qquad \text{for } x \in \Gamma_0,
	\label{cond_gamma0}
\ee
whereas the requirement of continuity in the field and its derivative at $\Gamma_1$ (i.e. the boundary conditions
at the interface) implies
\be
\begin{array}{rclr}
	\left. 
	\begin{aligned}
	\rho_0 u_0(x) - \rho_1 u_1(x) & = \; -\rho_0 \usrc(x)  \\
	\\
	\displaystyle \: \partial_n u_1(x) - \partial_n u_1(x) &= 
	 \: -\partial_n \usrc(x) \:
	\end{aligned} \right\} \; \text{for } x \in \Gamma_1
	\label{trans_cond}
\end{array}
\ee
The differential equations given in Eq. \eqref{helmh_eqs} and the conditions of Eqs. 
\eqref{cond_gamma0} and \eqref{trans_cond} constitute the boundary value problem 
that models the acoustic propagation problem schematized in Figure \ref{scheme1}.
In the next section this problem will be posed under a mathematical formulation appropriate
for an indirect boundary element method BEM.

\subsection{BEM formulation with a half-space Green function}
\label{formulation}


A typical indirect BEM formulation for the acoustic propagation problem previously introduced
presupposes that each field $u_i$ is expressed as a
linear combination of certain kind of surface integrals as, for example, in
\be
	u_i(x) =  a \int_\Gamma \Phi(x,y) \: \varphi(y) \: dS_y \: +  \:
		b \int_\Gamma \partial_{n_y} \Phi(x,y) \: \sigma(y) \: dS_y,
	\label{expansion_ejemplo}
\ee
where the {\it kernel} function $ \Phi(x,y) $ and its normal derivative are two-point functions
which give the field $x$-dependence while the integration (over the boundary $\Gamma$) 
is carried out in the $y$-variable, the latter emphasized by the sub-index in the surface differential $dS_y$.
Coefficients $a,b$ are known constants while $\varphi, \sigma$ are functions to be determined.
Each integral in Eq. \eqref{expansion_ejemplo} can be interpreted as an {\it integral operator} [Ref] 
acting on a function, for which the usual nomenclature is
\be
	U[ \varphi ](x) =  \int_\Gamma \Phi(x,y) \: \varphi(y) \: dS_y,
	\label{operator_U}
\ee
where $U$ represents the operator (associated with a kernel $\Phi$), and $\varphi$
the function which is integrated on the surface $\Gamma$.
As will be seen later, the field-values $\varphi(y)$ on $\Gamma$ will be unknown 
quantities for the integral formulation of our boundary value problem for acoustic propagation.  


In the vast majority of 3D BEM applications associated to the Helmholtz equation, the kernel function 
$ \Phi(x,y) $ is the free-space Green function in the wavenumber $ k_i $,
\be
	G(k_i;x,y) = \frac{ e^{ i k_i | x - y | } }{ 4 \pi| x - y | },
	\label{green}
\ee
which has the property of being a solution of the Helmholtz equation in absence of boundaries
(that is the reason why it is dubbed {\it free-space}).
It should be noted that if the kernel satisfies the Helmholtz equation (where the Laplacian
is taken with regard to the $x$-variable) then its associated operator also does it, as long as 
the interchange between the integral sign and the laplacian is possible.
This kernel $G(k_i;x,y)$ and its normal derivative define the operators 
\[
	S_i[ \varphi ](x) =  \int_\Gamma G(k_i;x,y) \: \varphi(y) \: dS_y,
	\qquad \qquad 
	K_i[ \varphi ](x) =  \int_\Gamma \partial_{n_y} G(k_i;x,y) \: \varphi(y) \: dS_y,
\]
in which the sub-index associates the operator with the wavenumber $k_i$ or equivalently with
the $R_i$ region.
These $S_i, K_i$ operators constitute the basis for all BEM methods and are called the single layer potential
operator (SLP) and the double layer potential operator (DLP), respectively.
Since the spherical symmetry of $G(k_i;x,y)$ offers no special advantages for the propagation problem
under consideration (because the boundaries are not spherical), a typical BEM formulation would 
involve integration in both $ \Gamma_0 $ and $ \Gamma_1 $, i.e. it would require numerical 
discretization of both these boundaries.

It is also worth noting that although the water and sediment layers are schematized in
Figure \ref{scheme1} without borderlines,  the $x,y$ directions, {in the underwater acoustic propagation modelling}
it is assumed that these limits are far away from the area of interest so their effects
can be neglected. In fact, this assumption implies that the solution for a limited area near
the source is identical to a solution for an acoustic problem which has those limits located at infinity.
By using this feature it is possible to use a modified Green function which takes into account 
part of the boundaries of the problem in an exact way, avoiding thus the need of building
a discretized version of those boundaries and consequently also avoiding to integrate over them.

\begin{figure}
    \centering
	\includegraphics[scale=0.5]{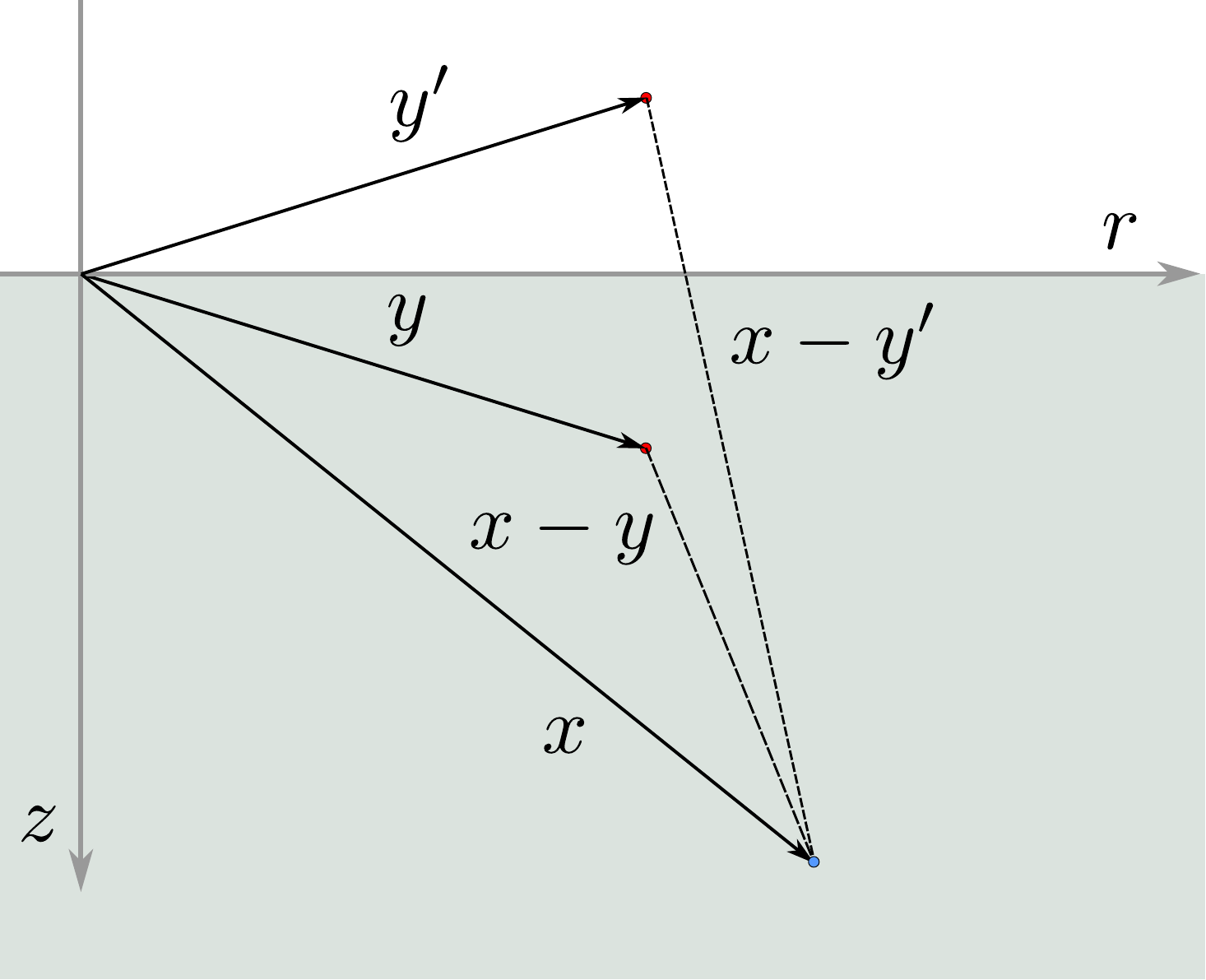}
	\caption{Points and vectors associated to the half-space Green function.
	For point $x,y$ appearing in the argument of the usual Green function, the half-space
	variant considers a point $y'$, specular reflection of $y$ by the $z=0$ plane.}
	\label{schemeGreen}
\end{figure}

Beyond the free-space Green function, the simplest useful modification is the half-space Green function,
obtained by using the {\it image method} \cite{jensen1994computational} which is of widespread application
in acoustics and optics (the Rayleigh-Sommerfeld correction to the problem of diffraction by an aperture being
an iconic example).
The half-space Green function appropriate to our problem will be the solution of the Helmholtz 
equation valid for a semi-space $z>0$ where the (infinite) boundary plane $z=0$ is such that over it 
the Green's function becomes zero. 
Its expression is
\be
	\HSgreen(k_i,x,y) = G(k_i;x,y) - \frac{ e^{ i k_i | x - y' | } }{ 4 \pi| x - y' | },
	\label{hsgreen}
\ee
where $y'$ is the specular image of the point $y$ by the plane $z=0$. 
The Figure \ref{schemeGreen} shows the points $x,y,y'$ involved in the definition of $\HSgreen$.
It is important to emphasize that the modification in Eq. \eqref{hsgreen} (an extra term) can be interpreted 
as a new source, an image source, located at point $y'$ outside of the physical region where the acoustic 
problem is being solved. 

{Regarding the propagation problem depicted in the Figure \ref{scheme1}, the use of a half-space
Green function as kernel in the $ S, K $ operators implies that the propagation problem satisfies
a pressure-release surface condition {\it built-in}, in the sense that $u_0$ from \eqref{expansion_ejemplo}
is equal to zero over $\Gamma_0$.
However, as it is stated in the condition \eqref{cond_gamma0}, it is desirable a litle more on $\Gamma_0$,
namely $u_0 + \usrc =0$, one way for that to happen is by redefining $u_0$ as $u_0 - \usrc $, where
$ \usrc $ is the image source}
\[
 	{\usrc}'(x) = \frac{1}{\rho_0} \frac{ e^{ i k_0 | x - {\xsrc}' |}}{ |x - {\xsrc}'| }.
\]

The problem solution in terms of these operators, plus the subtraction of the image source, automatically
will take into account the surface $\Gamma_0$ so that an explicit evaluation of the boundary conditions
there is not necessary. The fields $u_i$ have now to fulfill boundary conditions only at $\Gamma_1$.



Furthermore, writing the solution through the half-space Green function implies that the solution of the
propagation problem without the existence of $\Gamma_1$ (or with $\Gamma_1$ located at infinity) is
(or tends to) the image solution
\be
	\usrchsg = \usrc - {\usrc}',
\ee

\subsubsection{BEM formulation}

Now, after having chosen an appropriate kernel function, 
the fields $u$ in each region $R$ will have the following integral representation
\begin{equation}
	\begin{aligned}
	u_0(x) &=  S_{0}[\phi](x) + K_{0}[\psi](x) - {\usrc}'(x) \qquad \; \text{for } x \in R_0 
	\\
	\\
	u_1(x) &=  S_{1}[\phi](x) + K_{1}[\psi](x) \qquad \; \text{for } x \in  R_1
	\end{aligned}
	\label{potenciales}
\end{equation}
where 
\[
	S_{i}[\phi](x)=\int_{\Gamma_1} \HSgreen(k_i;x,y) \: \phi(y) \: dS_y
	\qquad \qquad
	K_{i}[\psi](x)=\int_{\Gamma_1} \partial_{n_y} \HSgreen(k_i;x,y) \: \psi(y) \: dS_y
\]
are the SLP and the DLP operators, evaluated on the unknown functions {$\phi$ and $\psi$, respectively}.
Since the boundary conditions of Eq. \eqref{trans_cond} involves the normal velocity it will be necessary 
to evaluate normal derivatives of these operators.
This procedure leads to two new operators,
\ben
	K_i'[\: \phi  \:](x) = 
	\partial_{n_x} \left( \int_{\Gamma_1} \HSgreen(k_i;x,y) \: \phi(y) \: dS_y \right) \qquad \qquad 
	T_i[\: \psi \:](x) = 
	\partial_{n_x} \left(\int_{\Gamma_1} \partial_{n_y} \HSgreen(k_i;x,y) \: \psi(y) \: dS_y\right),
\een
generically known as the \textit{normal derivative operators} \cite{Kress2001}.

The next step in the integral formulation is to evaluate the transmission conditions at $\Gamma_1$, Eq. 
\eqref{trans_cond}, using the field prescription according to \eqref{potenciales}.
This process requires to carefully consider the operator's {\it jump conditions} (section 3.1 of reference \cite{ColtonKress98}) or, in other words,
its behavior in the limit when its evaluation point $x$ in the surface is approached from $R_i$ along the 
normal vector direction $n_x$.
Following these steps the following system of two boundary integral equations
\begin{equation}
\begin{aligned}
	&\left.
	\begin{aligned}
	& ( \rho_0  K_0 - \rho_1  K_1 + 1/2 \: \alpha_{01}) \argpot{\psi}(x) + 
	(\rho_0  S_0 - \rho_1 S_1) \argpot{\phi}(x) = -\usrchsg(x) \; 
	\\
	\\
	& -\muller_{01} \argpot{\psi}(x) + (I - \rho_0 K'_0 + \rho_1 K'_1 )\argpot{\phi}(x)
	=  \partial_n\usrchsg(x)  \;
	\end{aligned} 
	\right\} \text{ for } x \in \Gamma_1 
	\label{masterBIE}
\end{aligned}
\end{equation}
for the two unknowns $\phi, \psi$ is obtained, where $\alpha_{01}= (\rho_0 + \rho_1) /2$ and 
$\muller$ is another operator, called the Müller operator and defined as
\be
	\muller_{s\ell}\argpot{\varphi}(x) \equiv T_s\argpot{\varphi}(x) -  T_\ell\argpot{\varphi}(x).
	\label{Muller_operator}
\ee
The Appendix \ref{App_integrals_from_bc} provides a schematic derivation of the system of 
Eq. \eqref{masterBIE} by evaluating carefully the limits and using the jump conditions from 
\cite{ColtonKress98}.

The modification in $\HSgreen$ with respect to the usual $G$, i.e. the addition of an image source at $y'$, 
modifies the usual expression for the integrands in the operators $S,K,K',T$. Nevertheless, these
modifications does not alter its continuity and jump properties because the added term in $\HSgreen$ is regular for all $x,y$ in the physical domain.
In the Appendix \ref{operators_hsgreen} expressions for all the four integral operator's integrands are provided.

The solution of the acoustic propagation problem given in terms of the fields $u_i$ has been transformed
in the search of the functions $ \psi, \phi $ {both having the
surface $\Gamma_1$ as their domains of definition}. These functions are called
{\it densities} in the literature and are now the unknowns. 
The BEM method provides a procedure to find discretized versions of these densities, which will be the subject
of the next section.

\subsection{Numerical method}
\label{numerical_method}

The system of Eq. \eqref{masterBIE} can be solved through a discretization process over the boundary $\Gamma_1$,
which turns it in a finite-size matrix system. 
For this step the standard procedure is to assume the following two approximations.
\begin{enumerate}
	\item  The surface $\Gamma_1$ is approximated by a planar triangular mesh (i.e., a set of $N_1$ triangles $\{ 
	\Delta_\ell \} $ with $\ell = 1,2,.. N_1$), so that
	\ben
		\Gamma_1 = \bigcup_{\ell=1}^{N_1} \Delta_\ell,
	\een
	where $\Delta_\ell$ is the $\ell$-th triangle whose centroid is $x_\ell$.
	\item	
	The unknown densities $\psi$ and $\phi$ are considered as piecewise constant functions in each triangle, 
	that is,
	\be
		\psi(x) = \sum_{\ell=1}^{N_1} \psi_\ell \: I_{\Delta_\ell}(x) \qquad \qquad 
		\phi(x)= \sum_{ \ell=1}^{N_1} \phi_\ell \: I_{\Delta_\ell}(x),
		\label{discr_fields}
	\ee
	where $\psi_\ell \equiv \psi(x_\ell), \phi_\ell \equiv \phi(x_\ell)$ are unknown complex numbers 
	and $ I_{\Delta_{\ell}}(x) $ is the indicator function of the $\ell$-th triangle, defined as 
	\[
		I_{\Delta_{\ell}}(x) = \begin{cases}
	                        1 \quad \text{ if } x \in \Delta_\ell \\
	                        0 \quad \text{ otherwise}
	                       \end{cases}
	\]
\end{enumerate} 

In order to find the densities, the prescriptions \eqref{discr_fields} are inserted in the system \eqref{masterBIE}. 
This procedure transforms each integral over the boundary $\Gamma_1$ into a sum of integrals over each triangle 
$\Delta_\ell$.

The resulting system remains, of course, valid for all $x \in \Gamma_1$, so that 
is {valid in particular} for the set of centroids $\{x_\ell\} (\ell=1,2,..,N_1)$ of the triangles belonging to $\Gamma_1$.
When these discretized equations are evaluated in the set $ \{x_\ell \}$, a matrix system of size 
$ 2 N_1 \times 2 N_1 $ 
\begin{equation}
	\left( \boldsymbol{B} + \boldsymbol{D} \right)
	\begin{pmatrix}	
	\boldsymbol{ \psi } \\
	\boldsymbol{ \phi }
	\end{pmatrix} =
	\begin{pmatrix}
	\boldsymbol{ f } \\
	\boldsymbol{ g }
	\end{pmatrix} 
	\label{sistema}
\end{equation}
is obtained. The complex-valued vectors $\boldsymbol{\psi}, \boldsymbol{\phi}$ containing the unknowns and the 
data vectors $\boldsymbol{f}$ and $\boldsymbol{g}$ are defined as
\begin{equation}
	\boldsymbol{ \psi } = \begin{pmatrix}
				\psi_1 \\
				\\
				\psi_2 \\
				\\
				\vdots \\
				\\
				\psi_{N_1}
				\end{pmatrix}
	\qquad
	\boldsymbol{ \phi } = \begin{pmatrix}
				\phi_1 \\
				\\
				\phi_2 \\
				\\
				\vdots \\
				\\
				\phi_{N_1}
				\end{pmatrix}
	\qquad
	\boldsymbol{ f } = -\begin{pmatrix}
				-\rho_0 \: \usrchsg(x_1) \\
				\\
				-\rho_0 \:\usrchsg(x_2) \\
				\\
				\vdots \\
				\\
				-\rho_0 \:\usrchsg(x_{N_1})
				\end{pmatrix}
	\quad 		
	\boldsymbol{ g } = \begin{pmatrix}
				\partial_n \usrchsg(x_1) \\
				\\
				\partial_n \usrchsg(x_2) \\
				\\
				\vdots \\
				\\
				\partial_n \usrchsg(x_{N_1})
				\end{pmatrix}.
	\quad 	
	\label{fgdef}	
\end{equation} 
The complex-valued matrices $\boldsymbol{B}$ and $\boldsymbol{D}$ are defined as $2\time2$ block-matrix, each block 
having size $N\times N$

%
%

%

\begin{equation} 
	\boldsymbol{B}= \left[
	\begin{array}{cc}
	\left(\rho_0 \boldsymbol{K_{0}} - \rho_1 \boldsymbol{K_{1}}\right) & 
	\left(\rho_0 \boldsymbol{S_{0}} - \rho_1 \boldsymbol{S_{1}}\right) 
	\\
	&
	\\ 
	- \boldsymbol\muller_{01} & (\boldsymbol{K_1'}-\boldsymbol{K_0'}) 
	\end{array}
	\right], 
\qquad 
	\boldsymbol{D}= \left[
	\begin{array}{cc}
		1/2 \: \alpha_{01} \boldsymbol{I} & \boldsymbol{0}
		   	\\
		\boldsymbol{0}&\boldsymbol{I} 
	\end{array}
	\right]
	\label{sistema2}
\end{equation} 
where $\boldsymbol{I}$ is the identity matrix of dimension $N_1$.
The boldface typography used for the operator's matrix representation attempts to capture its
discrete nature. For a generic operator $\boldsymbol{U}_q \in \mathbb{C}^{N_{1} \times N_{1}}$ 
with kernel $\Phi(k_q ; x, y )$ the $i,j$ matrix element represents
\[
	[ \boldsymbol{U}_q ]_{ij} = \int_{\Delta_j} \Phi(k_q; x_i,y) \: dS_y.
\]
The matrix row-index $i$ is associated with a particular evaluation point $x_i$ while the column index $j$ 
is associated with the particular element $\Delta_j$. 
The subindex $q$ identifies the corresponding wavenumber $k_q$.
For example, 
\[
	\left[ \boldsymbol{ S_{1} } \right]_{\ell s} =\int_{\Delta_s} \HSgreen_{k_1}(x_\ell,y) \: dS_y,
\]
implies integration over the $s$-th triangle of $\Gamma_1$ and evaluation on centroid $x_\ell$,
all for the wavenumber $k_1$.



%


Numerical evaluation of the operators can be cumbersome {since it} involves the resolution
of integral singularities. In this case, computer implementation is strongly based on the
previous work of \cite{gonzalez2020boundary} that uses the traditional free-space Green function
as integral kernel. Thus, the numerical evaluation of the matrix components at Eq.
\eqref{sistema2} 
is carried out by taking that previous implementation and changing the Green function.  
{Even though this procces is long and implies many calculations,} this is straightforward, because,
as was it previously stated, the kernel is modified by adding a term that is regular for all $x,y$
in the physical domain, meaning that there is no need to develop additional numerical treatments of singularities.

%

Once the matrix system of Eqs. \eqref{sistema}, \eqref{fgdef} and \eqref{sistema2} is solved and 
density vectors $\bfs{\phi}, \bfs{\psi}$ are obtained, we are ready to evaluate the field in any 
point $x$ belonging to the physical domain.

\subsection{Pressure and TL computation}
\label{tl_computation}

After having chosen the densities $ \bfs{\psi}, \bfs{\phi} $, the fields $u_i$
(in water or in the sediment layer) can be calculated by evaluating the
corresponding integral representation in Eq. \eqref{potenciales} with the piecewise approximation
made at Eq. \eqref{discr_fields}.
Therefore, using the transformation \eqref{field_transform} the discretized version of the pressure 
field in the water layer is 
\be
	p_0(x) = \:  - {\usrc}'(x) + \sum_{ \ell=1 }^{N_1}  
	\left( \rho_0 \: \phi_\ell 
	\int_{\Delta_\ell}\HSgreen(k_0;x,y) \: dS_y \: + 
	\rho_0 \: \psi_\ell 
	\int_{\Delta_\ell} \partial_{n_y} \HSgreen(k_0;x,y) \: dS_y \right).
	\label{nearfieldEq}
\ee

As it is common in acoustic propagation modelling, in the following sections results for 
acoustic propagation evaluation will be analyzed in the logarithmic scale using the
transmission loss (TL) parameter which is defined as  
\be
	\TL = -20 \log_{10} \left( \left| \frac{p_0(x)}{p_\text{ref}} \right| \right) \qquad \text{ dB re 1 m},
	\label{TL_def}
\ee
where the reference pressure $ p_\text{ref} $ is the pressure evaluated at 1 m from the source. 

\section{Model verifications: Pekeris waveguide}
\label{verification_pekeris}

In order to verify the BEM model the classical problem of propagation in a Pekeris waveguide 
was used as a benchmark solution.
It consists of a flat homogeneous water layer bounded below by a homogeneous half-space of higher
sound speed and above by a pressure-release interface. Such a setup, first analyzed within normal
mode theory \cite{pekeris1948theory}, is a simple but useful model for acoustic propagation in
shallow-water ocean environments.
This configuration is one of the most employed benchmark solutions because it allows an analytical
treatment by using wavenumber integration techniques and it is intrinsically important due to the
physical insight that it provides.
In the next subsection an overview of the exact solution and a benchmark case used for
verification is presented. Next, comparisons between BEM model and the exact solution are also included.

\subsection{Pekeris waveguide solution}

\begin{figure}[!h]
    \centering
	\includegraphics[scale=0.6]{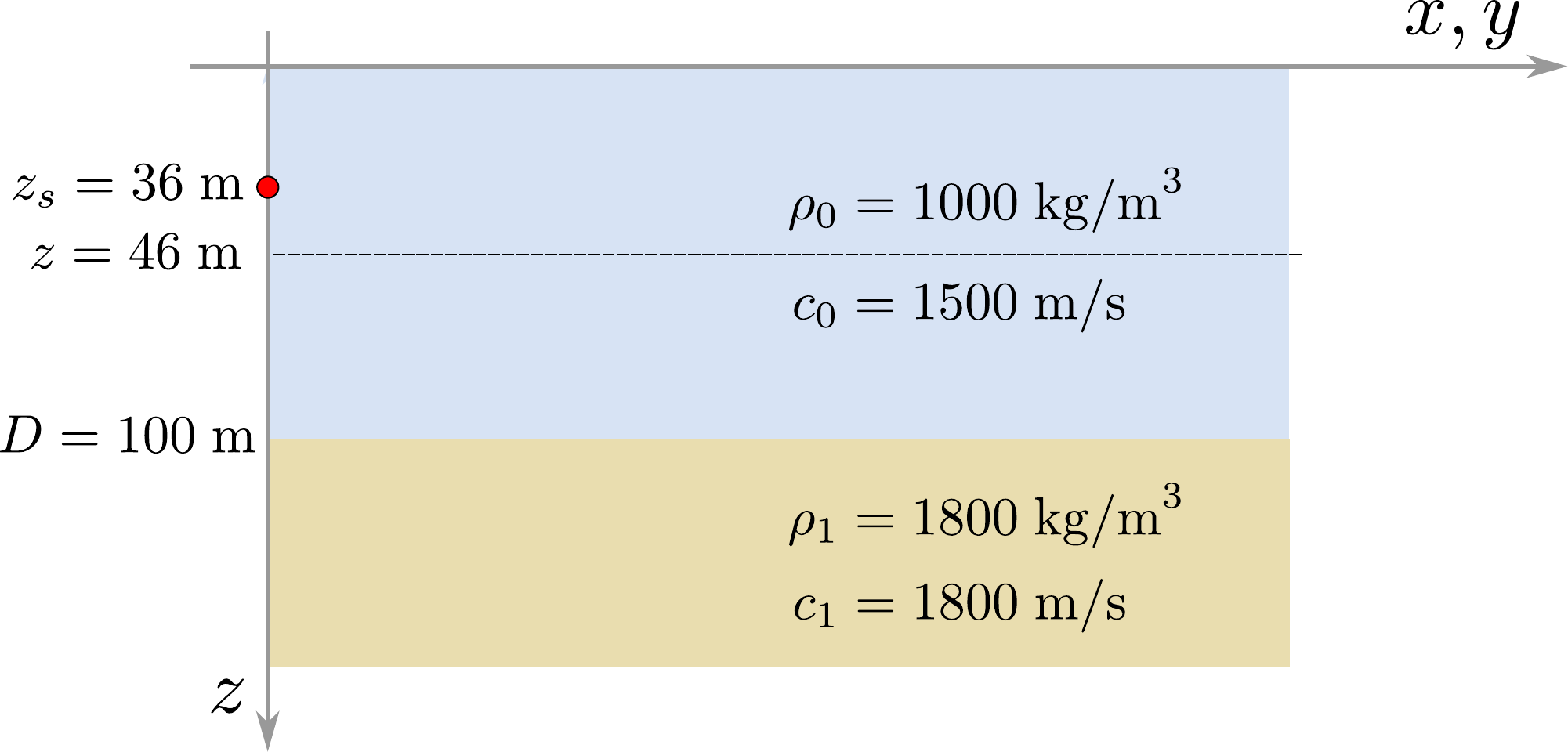}
	\caption{Pekeris waveguide of thickness $D$ = 100 m with pressure-release surface
	and penetrable bottom. The physical parameters $\rho,c$ in each layer are indicated
	in the figure.}
	\label{fig_pekeris}
\end{figure}

The Pekeris waveguide considered as our benchmark setup is shown in Figure \ref{fig_pekeris}.
A homogeneous water column of depth $D$, sound speed $c_0$ and density $\rho_0$ overlies an homogeneous
fluid half-space of sound speed $c_1$ and density $\rho_1$. 
A point source  located in water at $ (r,z) = (0,z_s) $  generates a pressure field 
\[
	p(r,z) = - \rho_0 \: \omega^2 \: S_\omega \frac{e^{ i k_0 |(r,z)-z_s|}}{4 \pi |(r,z)-z_s| },
\]
where $ S_\omega $ is the source strength.
The physical parameters $c,\rho$ and location $z_s$ of source and receiver depth $z$ used in this 
example were taken from an example in \cite{jensen1994computational} and are explicitly indicated
in the figure.

The pressure $p(r,z)$ in the Pekeris waveguide is then given by 
\[
	p(r,z) = \begin{cases}
		 \; \rho_0 \: \omega^2 \: \psi_0(r,z), \qquad 0 \leq z \leq D	\\ 
		 \; \rho_1 \: \omega^2 \: \psi_1(r,z), \qquad z \geq D	\\ 
	         \end{cases}
\]
where the $ \psi_i $ ($i=0,1$ corresponding to the water and sediments layers, respectively) are the displacement 
potentials which can be expressed as Hankel transforms
\[
	\psi_i(r,z) = \int_{0}^{\infty} \: \Psi_i(k_r,z) \: J_0( k_r r ) \: k_r \: dk_r,
\]
being $J_0$ the cylindrical Bessel function of first kind and $ \Psi_i(k_r,z) $ the solution of the depth-dependent 
wave equation
\[
	\left( \frac{d^2}{dz^2} + k_{z,i}^2 \right) \: \Psi_i(k_r,z) = \frac{S_\omega}{2\pi} \: \delta(z-z_s),
\]
where $ k_{z,i}^2 = k_i^2 - k_r^2 $ is the vertical wavenumber in the $i$-medium. 
Attenuation in the sediment layer is taken into account trough a complex wavenumber 
$ K = k_1 \: ( 1 + i \: \eta \: \alpha ) $,
where the coefficient $ \alpha $ quantifies its strength and $ \eta = 1 / ( 40 \:\pi \log_{10} e )$ is a numeric 
constant that turns dimensionless the imaginary part.
Usually, attenuation $\alpha$ is given in dB/wavelength.
Further details and physical analysis of this solution can be founded in \cite{jensen1994computational}.

\subsection{BEM model applied to the Pekeris waveguide}

The BEM model was compared to the Pekeris benchmark solution previously presented at frequency $f = 5$ Hz 
by evaluating transmission loss (TL) (a) along a 2500 m path at $ z = 46 $ m depth (inside the water
layer, see Figure \ref{fig_pekeris}) and 
(b) in a rectangle $2500 $ m width and $150$ m depth which encompass both water and sediment field
evaluations.
Two cases are considered; without attenuation and with an attenuation $\alpha = 5$ dB/$ \lambda$
(from the applications standpoint is a somewhat big value, its inclusion being justified only to test 
the model in a broad spectrum of its parameters).
All the simulations were conducted over a circular mesh of radius $ R = 2500 $ m and $N = 49009$ triangles 
which represents the boundless interface water-sediments.

\begin{figure}[!h]
    \centering
	\includegraphics[scale=0.5]{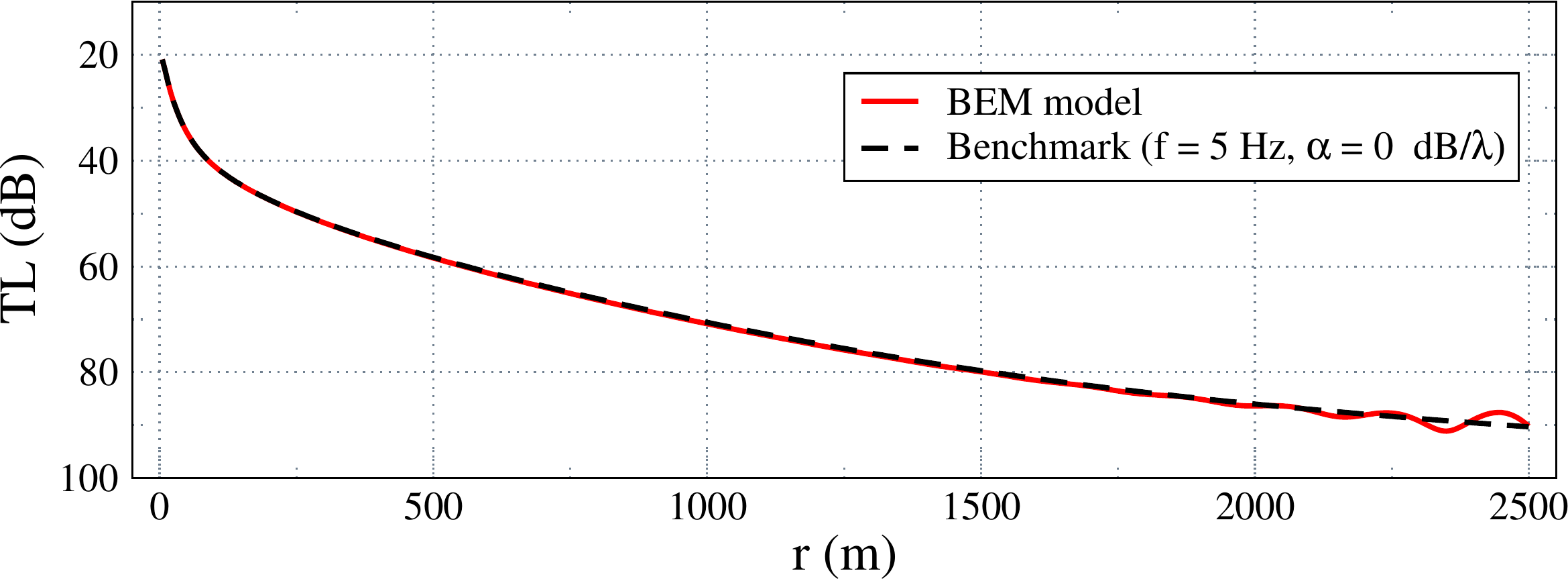}
	\includegraphics[scale=0.5]{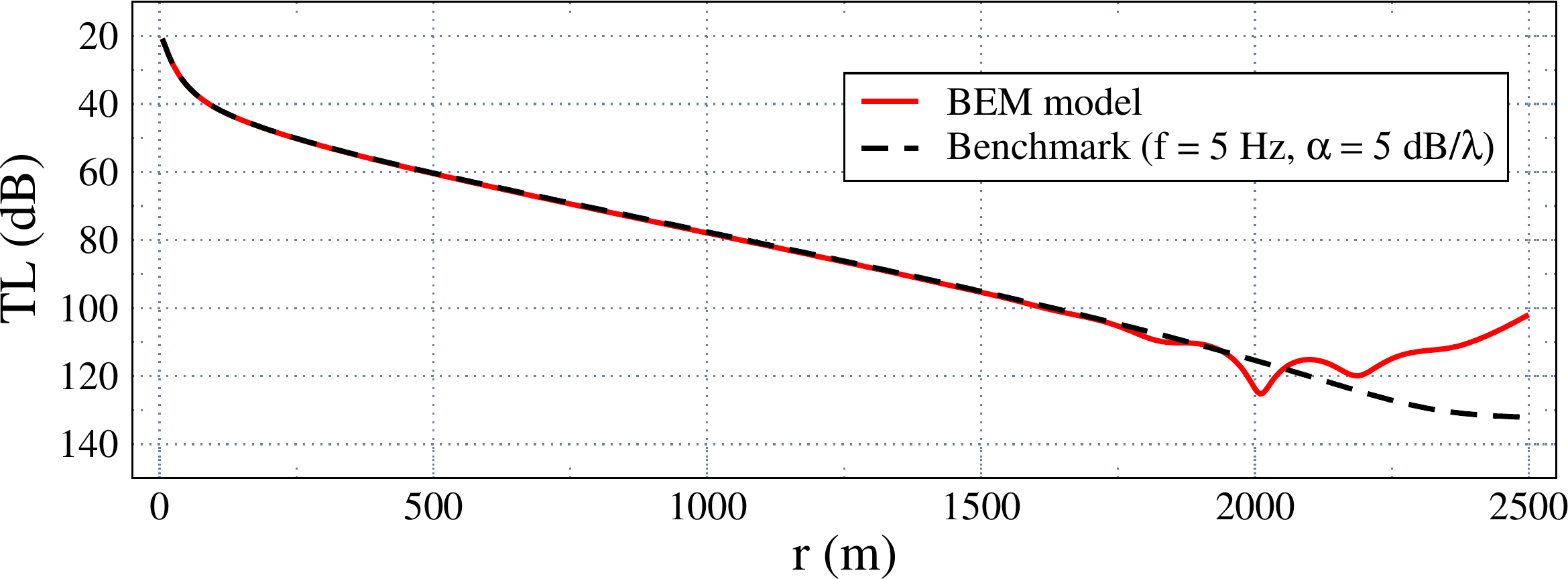}
	\caption{Transmission loss (TL) vs. range computed by the BEM model (full lines) and by the
	wavenumber benchmark solution (dashed lines) for a Pekeris waveguide with thickness of
	$100$ m (physical, source and evaluation parameters given in Figure \ref{fig_pekeris}) 
	along a 2.5 km path with depth $z=46$ m. Source frequency: $f = 5$ Hz.
	Top panel: without attenuation ($\alpha=0$ dB/$\lambda$). Bottom panel: with attenuation given by
	$\alpha = 5$ dB/$\lambda$.}
	\label{fig_pekeris_bench_1}
\end{figure} 

Since our BEM formulation assumes constant field on each triangle, the number $N$ of elements of the mesh 
must be chosen to ensure that the acoustic wavelength would be several times greater than the distance 
between vertices (usually five or six times greater) according to the usual practice in scattering applications
\cite{foote1980importance}.
In the circular mesh used for this verification test, the previous requirement is fulfilled by the 96 \% of the
triangle' sides.

\begin{figure}[!h]
    \centering
	\includegraphics[scale=0.35]{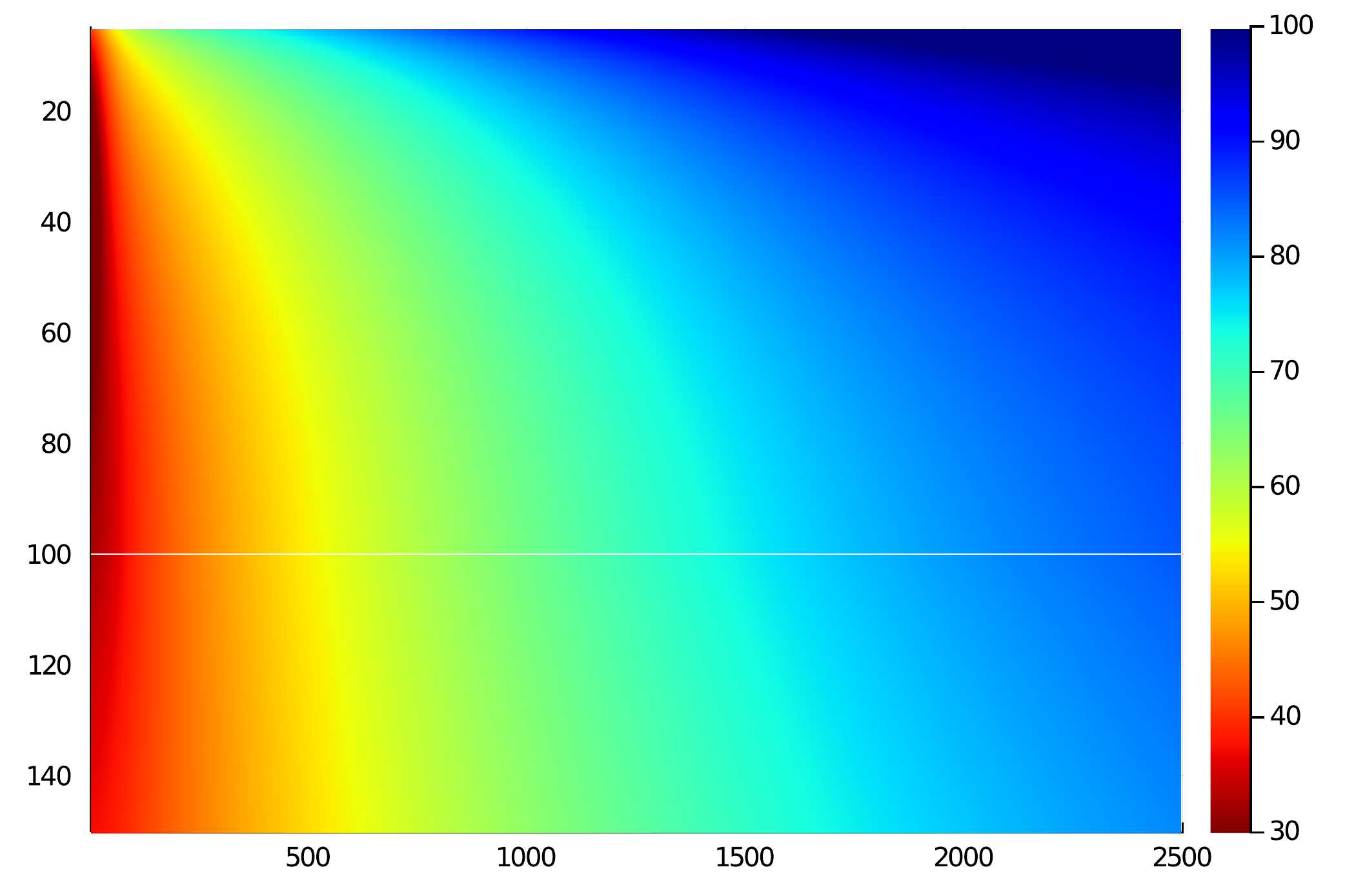}
	\includegraphics[scale=0.35]{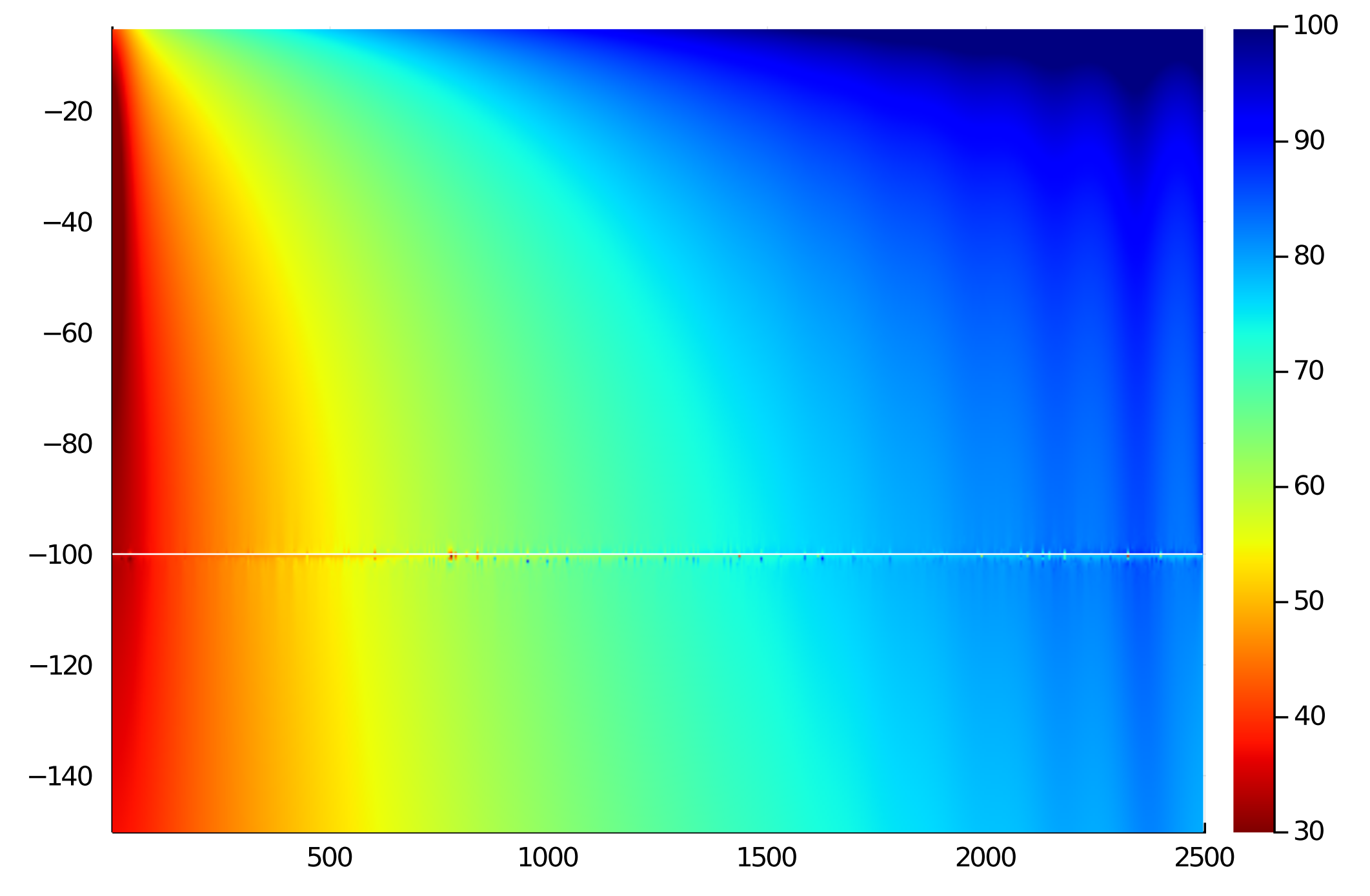}
	\caption{Transmission loss (TL) vs. range and depth, evaluated for the rectangle
	$0 \leq r \leq 2500$ m and $ 0 \leq z \leq 150 $ m (without attenuation). 
	Benchmark solution (left) and BEM solution (right) for $f = 5$ Hz. 
	The interface water-sediments is indicated by a white line.}
	\label{fig_pekeris_tl_rect_alpha_0}
\end{figure} 

The results of the comparison are shown in Figure \ref{fig_pekeris_bench_1}, where the top panel shows 
the case without attenuation ($ \alpha = 0 $) and the bottom one the case with an strong attenuation 
$ \alpha = 5$ dB/$\lambda $. 
The evaluation coordinate $r$ reachs the very end of the mesh so that errors are expected there. 
They are evident in the case $ \alpha \neq 0 $  but almost unnoticeable in the $ \alpha = 0$  dB/$\lambda $ case.
The former assumption about the pseudo-local character of the fields, i.e. the fact that the field 
at very long distances is not needed to determine the field in a restricted domain, appears to be a 
valid one. 
Of course, in cases where a benchmark solution is not available, 
which portions of the domain need to be meshed it should be object of analysis.

\begin{figure}[!h]
    \centering
	\includegraphics[scale=0.35]{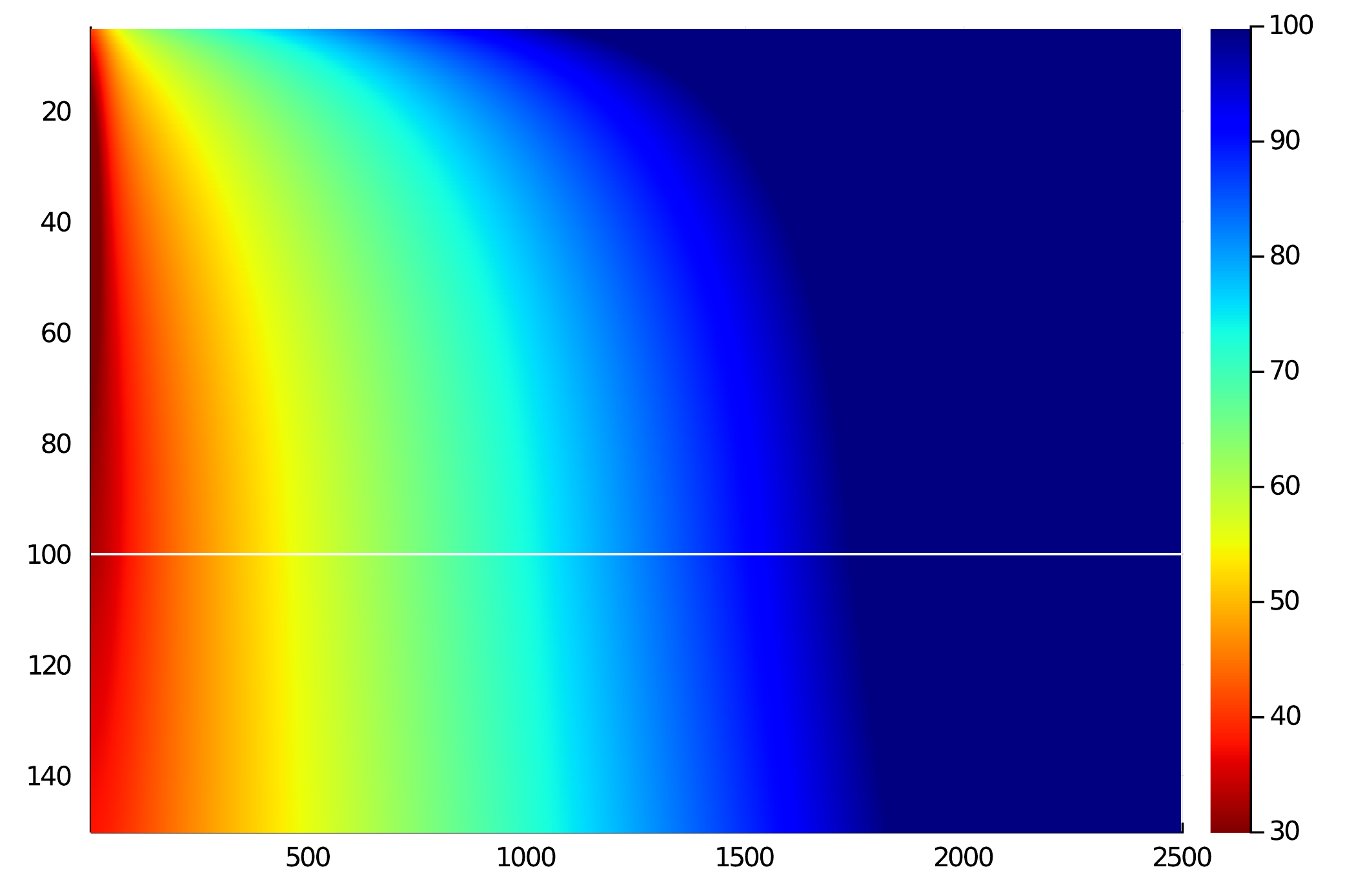}
	\includegraphics[scale=0.35]{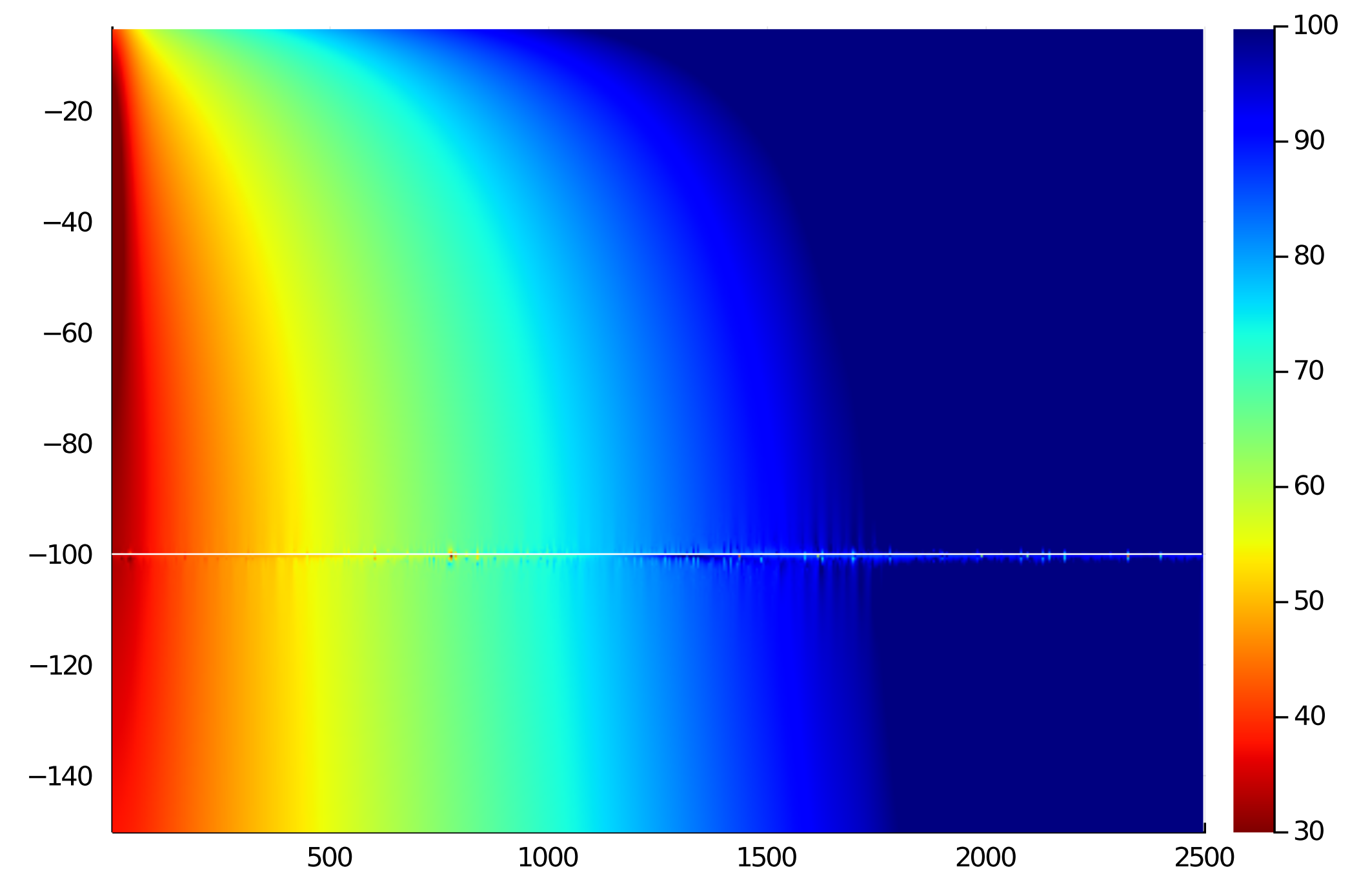}
	\caption{Transmission loss (TL) vs. range and depth evaluated for the rectangle
	$0 \leq r \leq 2500$ m and $ 0 \leq z \leq 150 $ m (with attenuation given by
	$\alpha = 5$ dB/$\lambda$). 
	Benchmark solution (left) and BEM solution (right) for $f = 5$ Hz. 
	The interface water-sediments is indicated by a white line.}
	\label{fig_pekeris_tl_rect_alpha_5}
\end{figure}

The Figures \ref{fig_pekeris_tl_rect_alpha_0} and \ref{fig_pekeris_tl_rect_alpha_5} show the
transmission losses for a rectangle which encompasses water and sediments in the cases of
$\alpha = 0$ and $\alpha = 5$ dB/$\lambda$ of attenuation, respectively. 
A solid white line indicates the interface between both media. The left panels show the benchmark solution 
whereas the right ones results provided by the BEM model. In all the figures colorbars
were saturated in the interval [30,100] dB for clarity purposes in the {comparison}.
The evaluations are visually in good agreement {although}  some numerical artifacts near the interface
line, which are a consequence of the evaluation point closeness in regard the triangle' sides in
the interface mesh. A finer mesh it will vanish this perturbations.

\subsection{Effect of the finiteness of the mesh}
\label{finiteness_mesh}

The propagation problem involves, as was previously mentioned, layers of almost unlimited extension.
The formulation based on the half-space Green function considers the pressure-release water surface
in an exact form, including also its intrinsic no-finiteness, but conversely the water-sediments
interface must be discretized --taking into account the relationship between wavelength and
segment length-- which arises a natural question: How far should this discretization go?


The previous verifications {\it a priori} assumed that the region surrounding
the source will be mandatory for an accurate representation of the field in the modelling,
therefore the source was centred regarding the disk constituting the mesh of the bottom surface.
To quantify the suitability of this centred-source assumption, we evaluate the TL corresponding
to the Pekeris waveguide (at $f = 5$ Hz and $\alpha = 0$ dB/$\lambda$) with the 
circular mesh of $2.5$ km radius and $N=49009$ triangles but shifting radially the source
location in 625, 1250 and 2500 m.
This provides four configurations which will be labelled $A, B, C, D$ being the first
them the original configuration (null shift).

\begin{figure}[!h]
    \centering
	\includegraphics[scale=0.5]{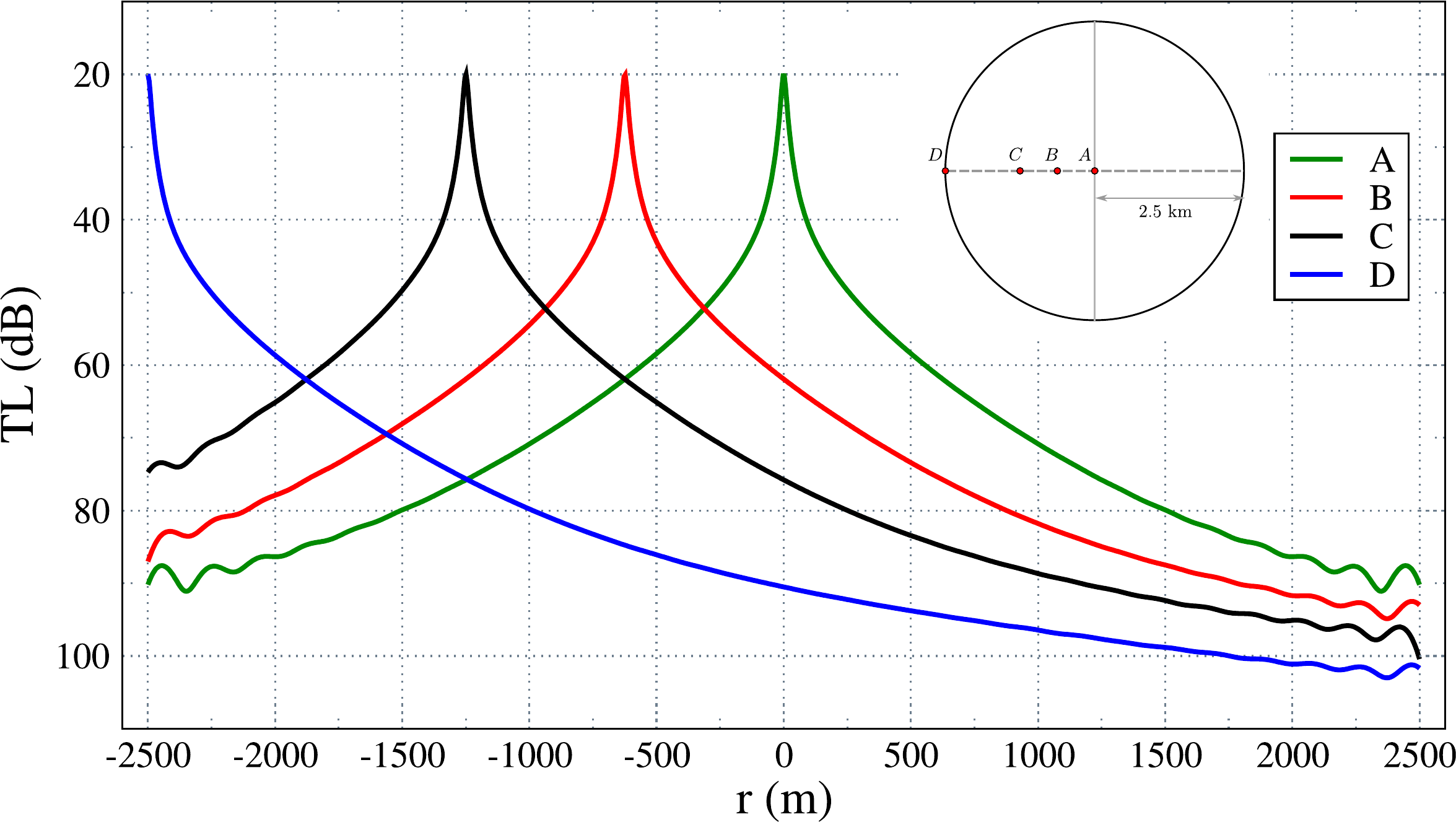}
	\caption{Transmission loss (TL) vs range computed by the BEM model for the Pekeris
	waveguide along a 5000 m radial path coincident with a radius of the circular
	mesh of $N = 49009$ triangles. The source were located at $z_s=36$ m but shifted
	radially in $r$ according to the scheme shown in the insert. The
	configurations $A, B, C $ and $D$ have the source at 0, -625, -1250 and -2500 in
	the coordinatization used.}
	\label{fig_pekeris_shifted_sources_circular}
\end{figure}

The resulting TL are shown in Figure \ref{fig_pekeris_shifted_sources_circular}
where the inset in the upper right corner identifies the source location in each case.
The source shift is indicated by a negative coordinate.
The finite size of the mesh is clearly noticeable at each side of the curves (except in the $D$ case) 
and the source location is evidenced by the sharp peak corresponding to minimum TL which
is obtained at the minimum distance ($z-z_s$) between source and evaluation point.

\begin{figure}[!h]
    \centering
	\includegraphics[scale=0.5]{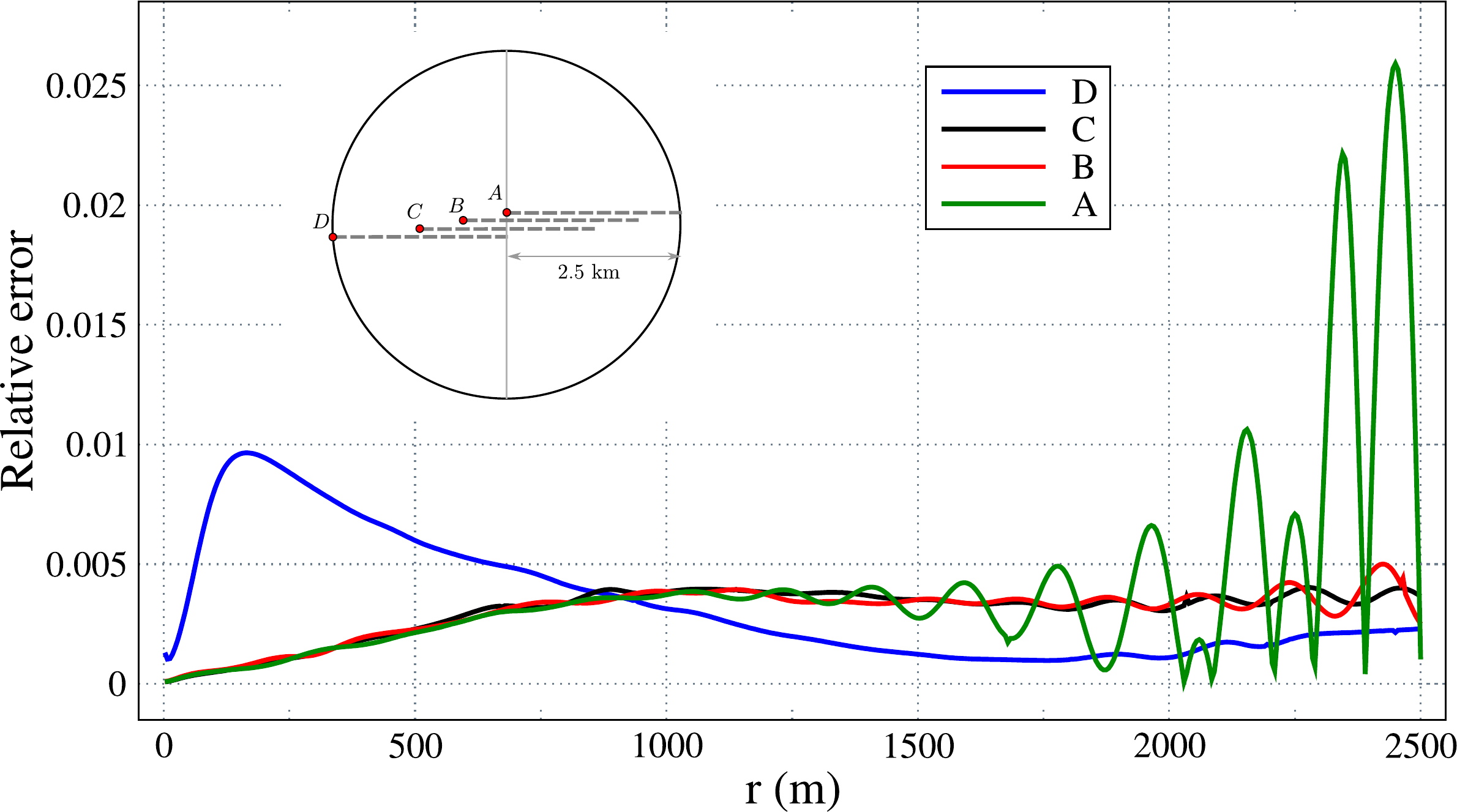}
	\caption{Relative error, according to definition \eqref{relerror}, for the TL vs range
	of shifted source configurations $A, B, C$ and $D$. In each case the first 2500 m from 
	the source to the right were compared point to point with the benchmark solution. 
	The inset in the upper left corner shows the evaluation segment compared for each
	case.}
	\label{fig_pekeris_error_mesh_circular}
\end{figure}

To quantify the error in each TL calculation, the resulting curves $B, C, D$ are shifted to match
with $A$. Then they are comparable point to point alongside a line of 2500 m, which is the minimum length
common to all of them. Defining a relative error in the logarithmic scale according to
\be
	 \text{Rel Err} = \frac{ \left| \TL - \TL_b \right| }{|\TL_b|}
	 \label{relerror}
\ee
where TL$_b$ is the Pekeris wavenumber integration prediction (the benchmark)
the relative errors associated to the different TL obtained in the four cases are shown
in Figure \ref{fig_pekeris_error_mesh_circular}. An inset in the left side shows a scheme for
source location in each configuration $A, B, C, D$ as well as the 2.5 km line segment (dashed lines)
where the TL are being calculated in each case (for clarity purposes -avoiding the overlapping- in this scheme
the lines are shifted a little amount in the vertical direction).

The $A$ configuration, i.e. the source ``centred'' at the mesh shows the expected error of mesh-finiteness,
already observed, clearly amplified towards the end of the line (beyond $r=2500$ m there is no world!).
Configurations $B$ and $C$, sources shifted in 625 and 1250 m, result in a better error behaviour,
because for neither one a relative error of 0.005 is surpassed. Remarkably, TL evaluation
for the immediate region surrounding the source is not perturbed in these shifted cases. 
The extreme case of a source located in the border of the domain, configuration $D$,
exhibits a maximum error in the first 250 m but a decreasing one afterwards.

These numerical experiments reveal that a compromise solution between accuracy and mesh use for TL 
calculation, along lines at least, can be attained considering a centered line of evaluation more
than a centered source. 
The associated question which arises next is, assuming that a centered configuration as $C$ satisfies
a desired error, how to surround that segment with a mesh big enough to provide an ``accurate'' TL
evaluation but simultaneously with little overhead due to calculation from far regions with negligible contribution.
To explore this question we considered two alternative meshes build from the original circular disk;
two ellipses of 2500 m major axis and minor axis in the relation 1:2 and 1:3, which results in 
1250 m and 833.33 m, respectively.

The two elliptical meshes were build assuring a triangle density similar to the one used in the 
circular mesh. Therefore the ellipses 1:2 and 1:3 have $N=24628$ and $N=16285$ triangles, respectively.
The relation between areas with respect to the 2500 m radius circular disk are 1/2 and 1/3 whereas
the relation between the number of triangles are 0.502 and 0.332. Therefore, the triangle density is
similar for all the three meshes.

\begin{figure}[!h]
    \centering
	\includegraphics[scale=0.50]{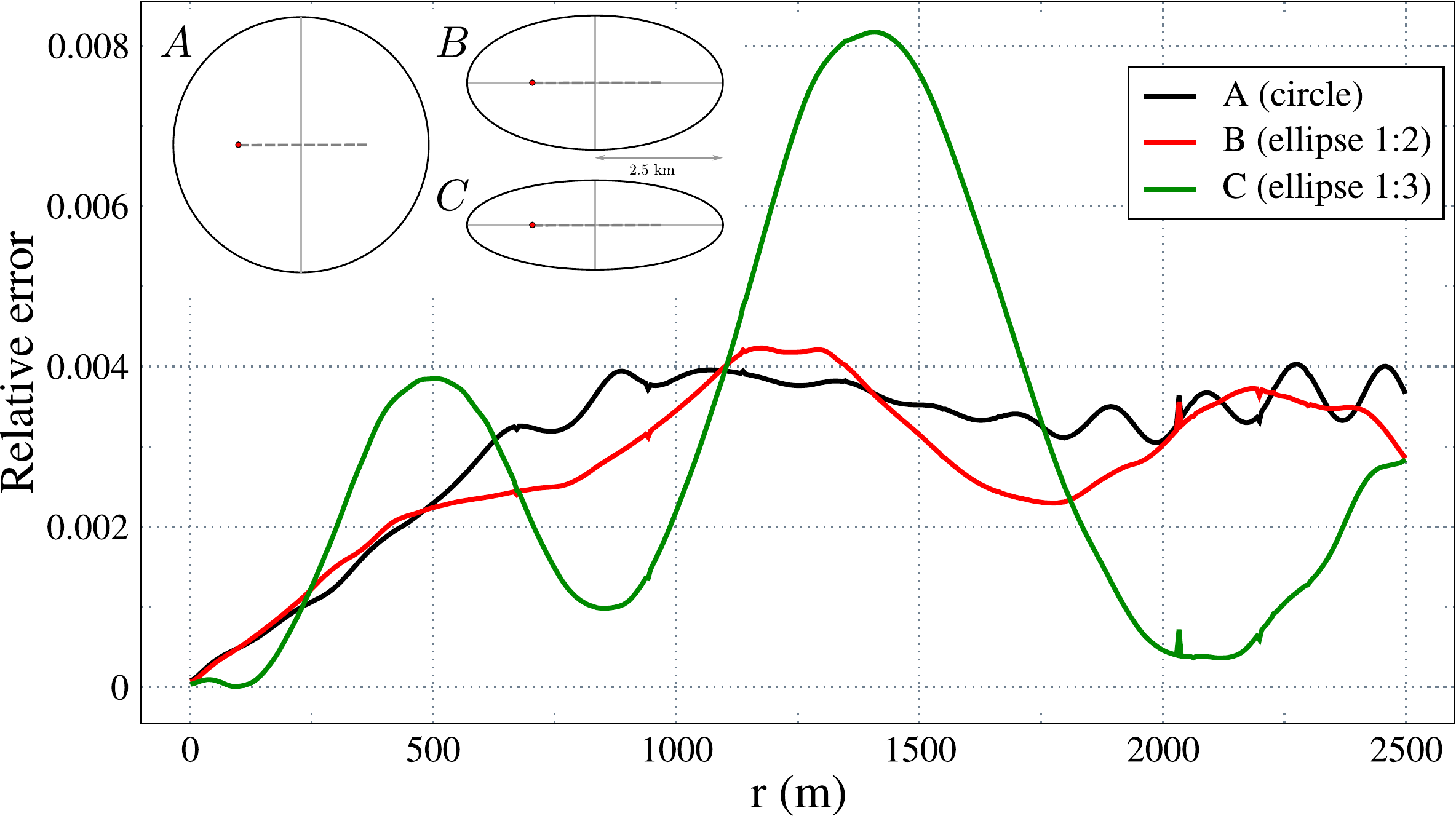}
	\caption{Relative error for TL vs range evaluated for a 2500 m segment, which is indicated by
	dashed lines in the scheme of the inset, for the 2500 m radius circular mesh and two elliptical meshes
	with major semiaxes of 2500 m and minor semiaxes in ratios 1:2 and 1:3.}
	\label{fig_error_meshes}
\end{figure}

In the Figure \ref{fig_error_meshes} the relative error for TL calculation along a 2500 m line,
is shown for the three meshes considered.
The inset at the upper left corner of the figure details the evaluation segment for the circular and
the elliptical meshes.
Surprisingly the decrease in the extension of the mesh has no appreciable effect in the TL since the 
relative error is similar.
The 1:3 ellipse displays an oscillatory behaviour, which emerges as an artifact introduced by the 
artificial elliptic boundary delimiting the end of the computational domain. 
Elliptical meshes with an increased number of triangles and other aspect-ratio as 1:4, display similar 
behaviors in $r$. Nevertheless, for the 1:3 ellipse, these relative errors do not exceed 0.64 dB
in absolute terms, a value which is acceptable in practical applications of acoustical propagation
in the ocean.

\section{BEM model in a shallow waters range dependent environment}
\label{bem_conical_mountain}

\begin{figure}[!h]
    \centering
	\includegraphics[scale=0.55]{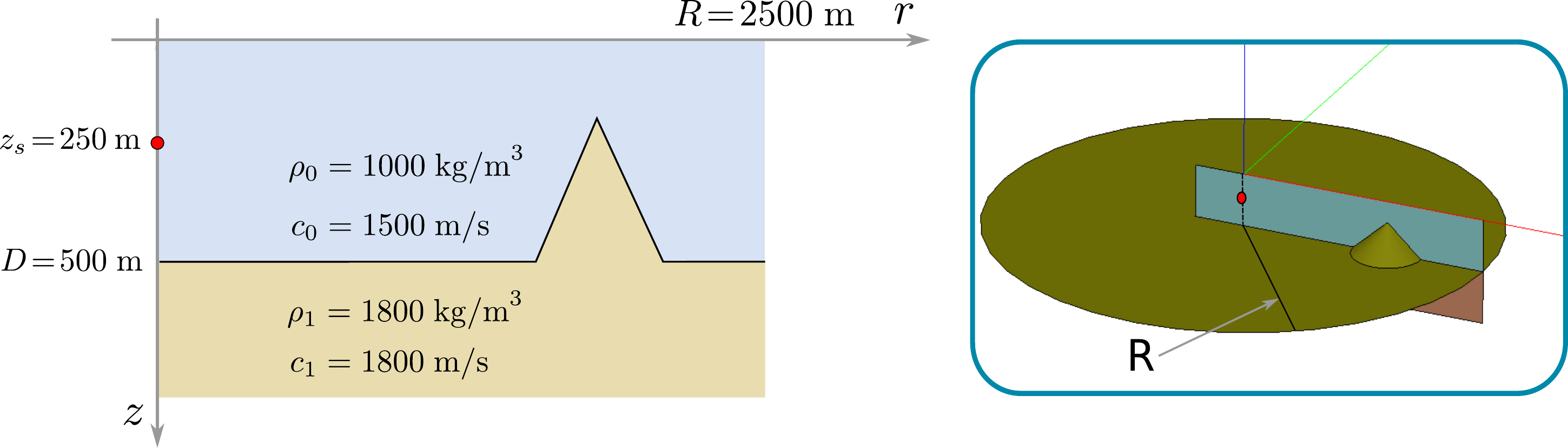}
	\caption{Scheme for a range dependent waveguide environment; a plane bottom  with a
	depth of 500 m provided with a conical mountain with a height of 300 m. TL will be evaluated over
	a plane of 2500 x 1000 m in $r,z$ which is shown with the propagation parameters 
	overimposed (left) and in a 3D panoramic view (right).}
	\label{fig_conical}
\end{figure}

To evaluate the model in a range dependent environment we have built
a synthetic scenario provided with a non-trivial bathymetry consisting of a plane bottom of
$ D = 500 $ m depth with a conical mountain with a height of $300$ m inside.
Similar scenarios were used, for example, in \cite{lin2013three}.
Transmission losses will be evaluated over a $r,z$ plane of 2500 x 1000 m which cuts in half the mountain.
A point source of unitary amplitude and frequency $f = 5$ Hz is located at $r=0$ and $z_s=250$ m.
A mesh of circular shape with radius $ R = 2500 $ m containing the mountain surface will be used.
The Figure \ref{fig_conical} shows a 2D sketch of the evaluation plane with the physical 
parameters indicated over it (left) and a 3D view encompassing both, the plane and the water-sediments
interface mesh (right).

The circular mesh has $ N = 50000 $ triangles. 
The given frequency involves acoustics wavelengths of $\lambda_0 = 300$ m and $ \lambda_1 = 360 $ m 
for the water and sediments, respectively. Considering a maximum segment length
of $\lambda/5$, and taking into account the lower wavelength it leads to a triangle' sides $\ell < 60$ m.
This condition was fulfilled for the 95 \% of the triangles of the mesh.

\begin{figure}[!h]
    \centering
	\includegraphics[scale=0.47]{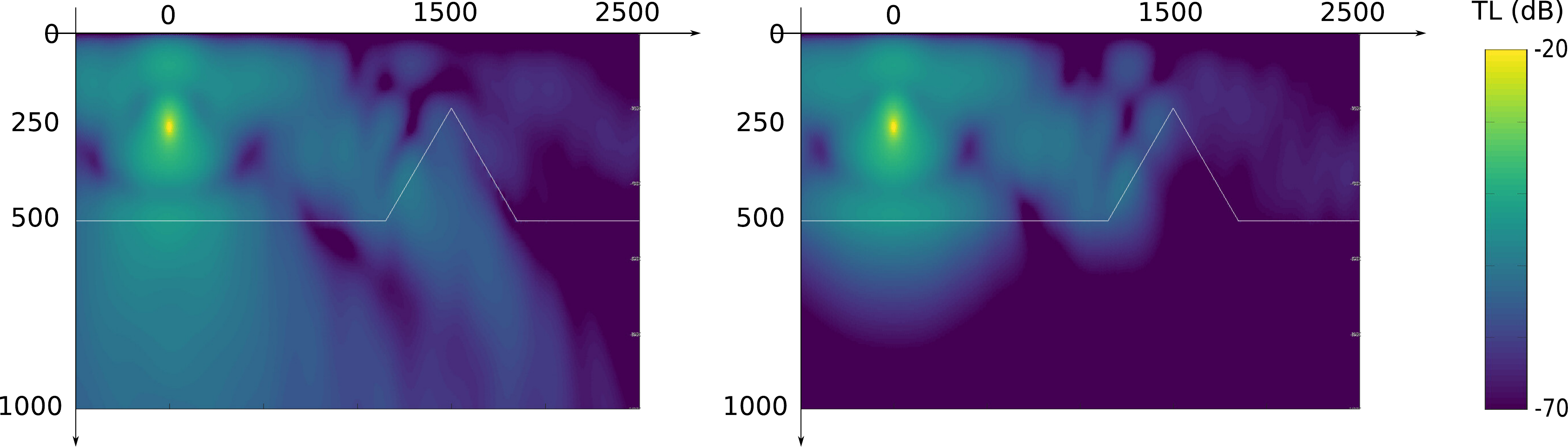}
	\caption{Transmission loss (TL) evaluated over the $(r,z)$-plane previously shown 
	in the Figure \ref{fig_conical}.
	The white line indicates the water-sediments interface. The left panel considers attenuation $\alpha=0$
	whereas the right one $\alpha=20$ dB/$\lambda$. The color bar on the right is saturated for 
	a better contrast.}
	\label{fig_conical_tl_bem}
\end{figure}

For TL evaluation two attenuation coefficients were used; $\alpha = 0$ (no attenuation) and $\alpha = 20$
dB/$\lambda$, the latter a value big enough only aimed to show a remarkably different graphical result.
In Figure \ref{fig_conical_tl_bem} results for TL in the $rz$-plane are shown: the no attenuation case
(left panel) and the $\alpha = 20$ dB/$\lambda$ one (right panel). 
The color bar values were saturated for a better contrast.
The water-sediments interface is indicated by a white line showing that the TL changes continuously 
when the interface is crossed. The region surrounding the source shows brighter indicating low TL values,
as were expected.

The intense attenuation due to the selected $\alpha$-value accounts for the fast vanishing
of the pressure field and consequent increase of the transmission losses. 
Moreover, it is noticeable that the conical mountain presence does not preclude appreciably pressure
fields behind it; which is an effect of the relative low frequency and diffraction effects due to the boundaries.

\section{Conclusions}
\label{conclusions}

For an underwater  acoustic propagation problem the BEM approach provides a clean formulation
regarding the field behaviour at infinity, as a consequence of automatically fulfilling of the radiation condition.
There is no need of artificial boundaries and layers, required in other methods to avoid emergence 
of spurious numerical artifacts.
Discretization of the entire domain boundary is not required to achieve an acceptable prediction but still
this ``local approach'' to the problem can still be computationally demanding at the frequency ranges of
interest for some applications.
The formulation based on a half-space Green function requires some tweaking of the integral 
operators for enabling to consider an infinite pressure release surface (Dirichlet condition).
An infinite acoustic hard surface (Neumann condition) suitable for modelling propagation over 
a rigid bottom can be worked out in the same fashion with minor modifications.

For an admissible accuracy in the TL evaluation a mesh considering a limited region near the source
is enough.
Even for moderated frequencies that implies very large matrix systems; thus, there is a clear need of
iterative methods for the system solving stage and accelerated integral evaluations, two key components
to avoid the explicit matrix construction and fast evaluation of the matrix elements, respectively.

\section*{Author Contribution and Funding} 
EL: algorithm implementation, theoretical development, numerical simulations, methodology, formal analysis,
and writing – original draft. 
JDG: theoretical development, methodology, formal analysis, and writing – original draft.
SB: methodology, formal analysis, and writing – review and editing. 
All authors contributed to the article and approved the submitted version. 
This work was supported by the Program of the Argentinian Ministry of Defense (PIDDEF 02/20), the Argentinian Navy and the National Council for Scientific and Technical Research (CONICET)

\appendix
\section{ Boundary integral equation from boundary conditions at $\Gamma_1$}
\label{App_integrals_from_bc}

\begin{figure}[!h]
    \centering
	\includegraphics[scale=0.5]{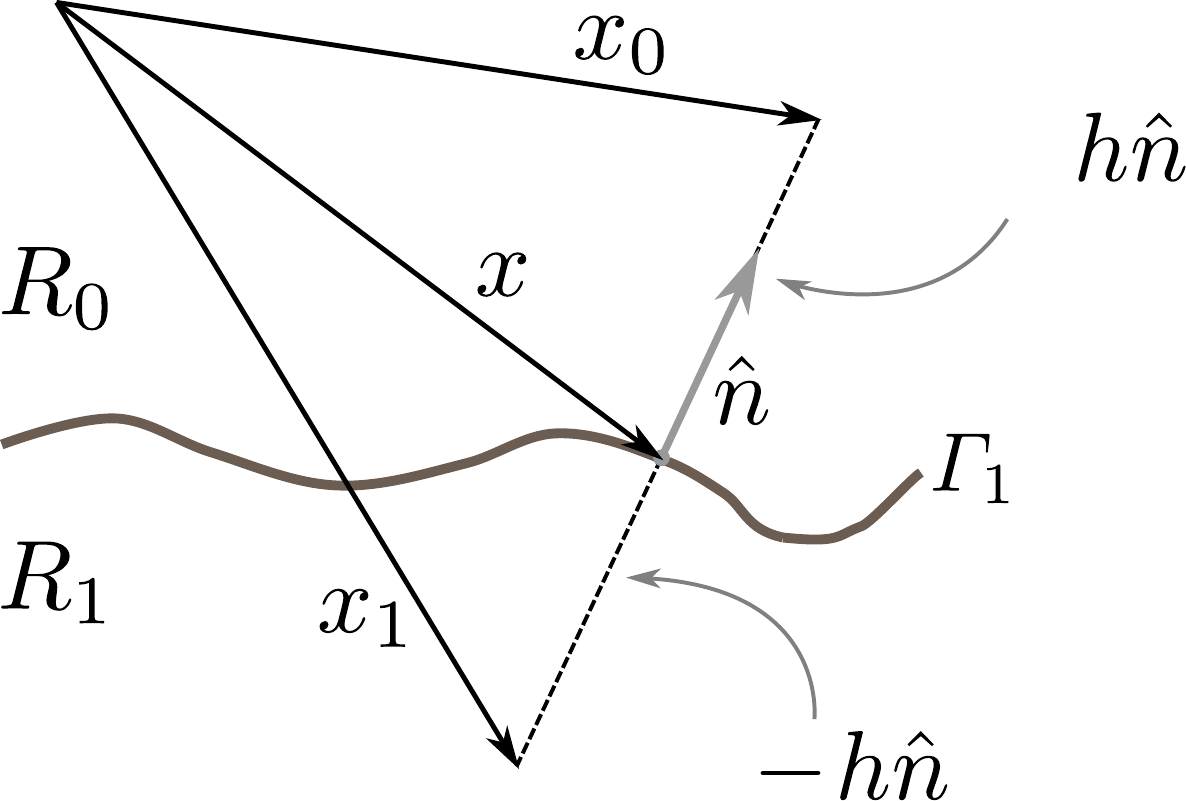}
	\caption{Sketch for normal derivative evaluation.}
	\label{fig_derivada_normal}
\end{figure}

Evaluation of an integral operator $U[\varphi]$ at a boundary point $x \in \Gamma_1$  
approached from the region $R_0$, is calculated as the limit
\[
	U[\varphi](x) = \lim_{h \to 0} \: U[ \varphi ]( x + h \hat{n} ),
\]
where $x$ is reached along the normal direction given by $\hat{n}$ (the normal at $x$)
from the point $x_0 \in R_0$ (see Figure \ref{fig_derivada_normal}). 
The operator evaluation at $x$ but approached from the region $R_1$ proceed along the
opposite direction (i.e. along $-\hat{n}$) from the point $x_1 \in R_1$, i.e. 

\[
	U[\varphi](x) = \lim_{h \to 0} \: U[ \varphi ]( x - h \hat{n} ),
\]
where $h>0$ in both cases.

By inserting the prescriptions \eqref{potenciales} in Eqs. \eqref{trans_cond} and defining
$ x^+ = x + h \hat{n} $  and $ x^- = x -h \hat{n}$ the boundary conditions can be
considered as $h \to 0$ limits, 
\[
	\lim_{h\to 0} \quad 
	\rho_0 \: ( \: S_0[\phi](x^+) + K_0[\psi](x^+) + \usrchsg(x^+) \: ) =
	\lim_{h\to 0} \quad
	\rho_1 \: ( \: S_1[\phi](x^-) + K_1[\psi](x^-) \: )
\]
\[
	\lim_{h\to 0} \quad 
	K_0'[\phi](x^+) + T_0[\psi](x^+) + \partial_n \usrchsg(x^+) =
	\lim_{h\to 0} \quad 
	K_1'[\phi](x^-) + T_1[\psi](x^-)
\]
where the evaluation point $x^+$ tends to $\Gamma_1$ from $R_0$ whereas point $x^-$ tends from $R_1$.
Evaluating the $h\to 0$ limit and considering the jump (\textit{jump and continuity conditions},
see theorem 3.1 of \cite{ColtonKress98}) for $K_i$ and $S_i$, we obtain
\[
	\rho_0 \: S_0[\phi](x) - \rho_1 \: S_1[\phi](x) + 
	\rho_0 \: K_0[\psi](x) - \rho_1 \: K_1[\psi](x) + \alpha_{01} \: I [\psi](x) =
	-\rho_0 \: \usrchsg(x)
\]
\[ 
	K_0'[\phi](x) - K_1'[\phi](x) -I [\phi](x) + 
	T_0[\psi](x) - T_1[\psi](x) =
	- \partial_n \usrchsg(x)
\]
where $\alpha_{01}= (\rho_0+\rho_1)/2$ and $I$ is the identity operator.
Using Müller operator' definition, Eq. \eqref{Muller_operator}, and a more compact notation 
the boundary conditions results in 
\[
	( \: \rho_0 \: S_0 - \rho_1 \: S_1 \: ) [\phi](x) + 
	( \: \rho_0 \: K_0 - \rho_1 \: K_1 + \alpha_{01} \: I ) [\psi](x) =
	-\rho_0 \: \usrchsg(x)
\]
\[ 
	( \: K_0' - K_1' - I \: ) [\phi](x) + \muller_{01}[\psi](x) = - \partial_n \usrchsg(x).
\]

Multiplying the last equation by -1, an usual convention, it results in the boundary integral
system of Eq. \eqref{masterBIE}.

\section{Operator's expressions for $\HSgreen$}
\label{operators_hsgreen}

\begin{figure}[!h]
    \centering
	\includegraphics[scale=0.5]{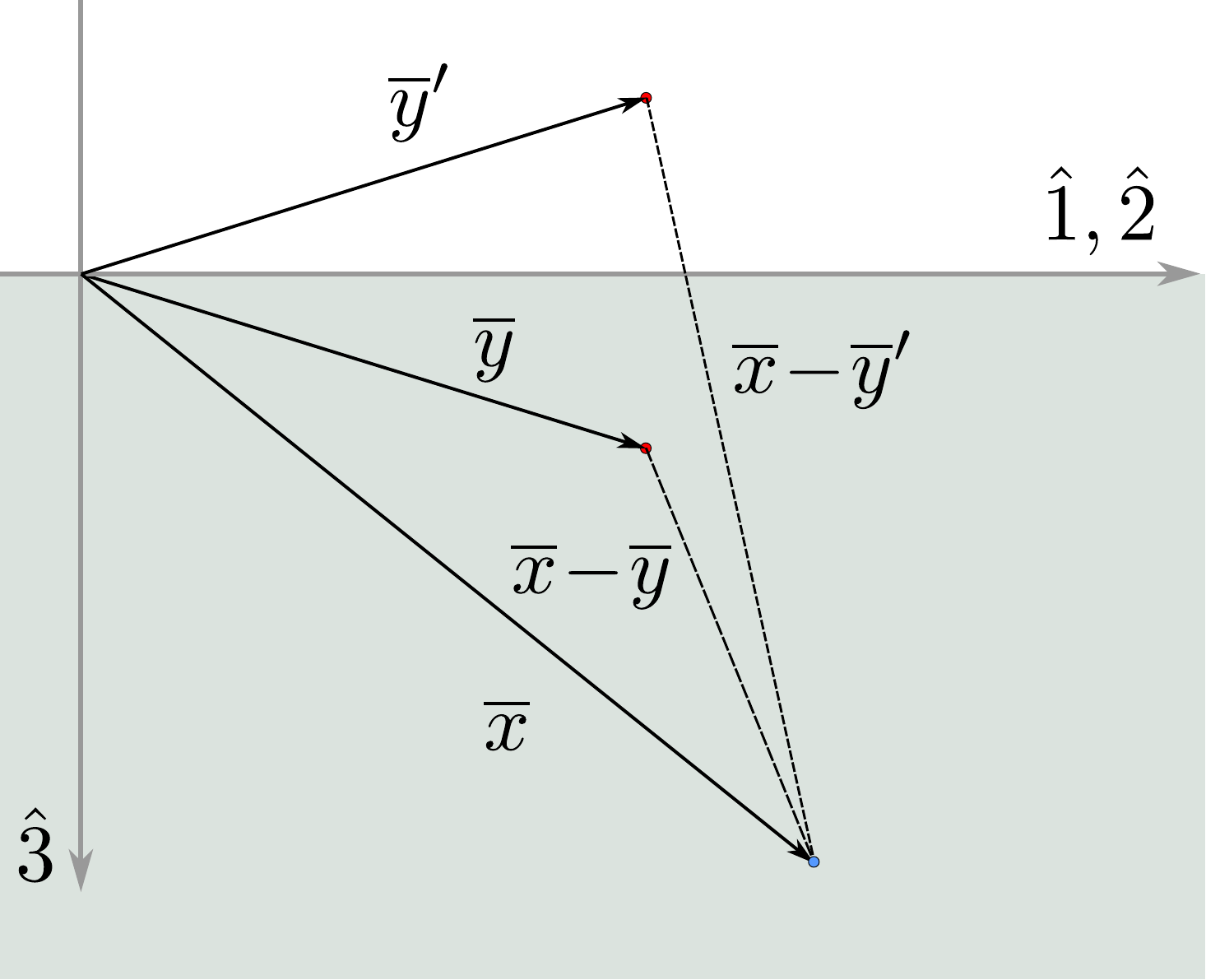}
	\caption{Coordinate system used for the graphical representation of vectors
	and distances involved in the half-space Green function operator's expressions.}
	\label{green_scheme_app}
\end{figure}

The cartesian coordinate system associated to the problem is shown in Figure \ref{green_scheme_app}.

Throughout this appendix the vector property will be explicitly emphasized by a line over the 
corresponding letter, for clarity reasons and to avoid confusions. 
With this convention the points under consideration are 
$\vet{x} = (x_1,x_2,x_3), \: \vet{y} = (y_1,y_2,y_3)$ and $\vet{y}' = (y_1,y_2,-y_3)$,
being useful to define 
\[
	r = |\vet{x} - \vet{y}|, \qquad \qquad r' = |\vet{x} - \vet{y}'|
\]
and unitary vectors
\[
	\hat{r} = \frac{\vet{x} - \vet{y}}{r} = \frac{ \vet{r} }{ r }, 
	\qquad \qquad
	\hat{r}' = \frac{\vet{x} - \vet{y}'}{r'} = \frac{ \vet{r}' }{ r' }.
\]

The half-space Green function in the wavenumber $k$ is
\be
	\HSgreen(k,\vet{x},\vet{y}) = \frac{ e^{ i k | \vet{x} - \vet{y} | } }{ 4 \pi| \vet{x} - \vet{y} | } - 
	\frac{ e^{ i k | \vet{x} - \vet{y}' | } }{ 4 \pi| \vet{x} - \vet{y}' | } =
	\frac{ e^{ i k r } }{ 4 \pi r } - \frac{ e^{ i k r' } }{ 4 \pi r' }
	\label{hsgreen_app}
\ee

Given that the Green function is a two-point function, it is relevant to indicate respect to what point 
the gradient is considered. It will be noted with a subscript in the nabla symbol; $\nabla_x$ or $\nabla_y$.
Since the dependence of the coordinates in $\HSgreen$ is mediated by $r$ and $r'$ it is convenient 
to use the chain rule $ \nabla f(r) \equiv \partial_r f \: \nabla r $ and 
$ \nabla f(r') \equiv \partial_{r'} f \: \nabla r' $ in the gradient calculation. 
By using this idea and the relations
\[
	\nabla_x \: r  = \frac{\vet{r}}{r}, \qquad\quad
	\nabla_y \: r  = -\frac{\vet{r}}{r}, \qquad\quad 
	\nabla_x \: r'  = \frac{\vet{r}'}{r'}, \qquad\quad 
	\nabla_y \: r'  = -\frac{1}{r'} \: \breve{r}'
\]
where $\: \breve{r}' \:$ is a vector that results from multiplying $\vet{r}'$ by a diagonal matrix which 
has its third coordinate negative, i.e.
\[
	\breve{ r }' = \begin{pmatrix}
			1 & 0 & 0 \\
			0 & 1 & 0 \\
			0 & 0 & -1
		    \end{pmatrix}
		    \: \vet{ r }',
\]
the following useful results are obtained
\[
	\nabla_x\left( \frac{ e^{ i k r } }{ 4 \pi r }  \right) = 
	\frac{ e^{ i k r } }{ 4 \pi r^3 } \: \left( i k r - 1 \right) \: \vet{r}, 
	\qquad 
	\nabla_x\left( \frac{ e^{ i k r' } }{ 4 \pi r' }  \right) = 
	\frac{ e^{ i k r } }{ 4 \pi r^{'3} } \: \left( i k r' - 1 \right) \: \vet{r}',
\]
\[
	\nabla_y\left( \frac{ e^{ i k r } }{ 4 \pi r }  \right) = 
	-\frac{ e^{ i k r } }{ 4 \pi r^3 } \: \left( i k r - 1 \right) \: \vet{r}, 
	\qquad 
	\nabla_y\left( \frac{ e^{ i k r' } }{ 4 \pi r' }  \right) = 
	-\frac{ e^{ i k r } }{ 4 \pi r^{'3} } \: \left( i k r' - 1 \right) \:
	\breve{r}'.
\]

With these ingredients the kernels of the four operators are
\begin{itemize}
\item For $S[\: \varphi \:](x)$ 
\[
\HSgreen (x,y) =  \frac{1}{4\pi} \left[ \: \frac{ i k r }{ r } - \frac{ i k r' }{ r' } \: \right]
\] 


\item For $K[\: \varphi \:](x)$ 
\[
\partial_{n_y} \HSgreen(x,y) = 	\frac{1}{4\pi} \left[ \: \frac{ i k r }{ r^2 }( ikr -1)(-\hat{r}\cdot \hat{n}_y) - 
	\frac{ i k r' }{ {r'}^2 }(i kr' -1) \left( -\frac{\breve{r}'}{r'} \cdot \hat{n}_y \right) \: \right]
\] 


\item For $K'[\: \varphi \:](x)$
\[
	\partial_{n_x} \HSgreen \: dS_y =	\frac{1}{4\pi} \left[ \: \frac{ i k r }{ r^2 }( ikr -1)(\hat{r}\cdot \hat{n}_x) - 	\frac{ i k r' }{ {r'}^2 }(i kr' -1) \left( \hat{r}' \cdot \hat{n}_x \right) \: \right]
\]


\item For $T[\: \varphi \:](x)$
\begin{multline*}
	\partial_{n_x} \partial_{n_y} \HSgreen_k(x,y) = \\ 
	\frac{ e^{ i k r } }{ 4 \pi r^3 } \left\{ 
	[ 3 ( i k r - 1) + k^2 r^2 ] \: ( \hat{r} \cdot \hat{n}_x )(\hat{r} \cdot \hat{n}_y) -
	( i k r - 1 ) \: \hat{n}_x \cdot \hat{n}_y \right\} + \\
	\frac{ e^{ i k {r'} } }{ 4 \pi {r'}^3 } \left\{ 
	[ \: 3 ( i k r' - 1) + k^2 {r'}^{2} \: ] \: 
	( \hat{r} \cdot \hat{n}_x )\left(\frac{\breve{r}'}{r'} \cdot \hat{n}_y\right) - 
	( i k r' - 1 ) \: \partial_{n_x}( \breve{r}' \cdot \hat{n}_y ) \right\}, 
\end{multline*}
where the last factor in the second term of the $T$ operator is 
\[
	\partial_{n_x}( \breve{r}' \cdot \hat{n}_y ) =
	\hat{r}'\cdot \hat{n}_x \: \left( \frac{\breve{r}'}{r'} \right)\cdot \hat{n}_y
	\; + \; 
	\hat{n}_x \cdot \begin{pmatrix}
	            1 & 0 & 0 \\
	            0 & 1 & 0 \\
	            0 & 0 & -1
	            \end{pmatrix} \hat{n}_y \; - \; \hat{n}_x \cdot \left( \frac{\breve{r}'}{r'} \right)
	            \hat{r}' \cdot \hat{n}_y .
\]

\end{itemize}




\end{document}